\documentclass[11pt,a4paper]{article}
\usepackage[utf8]{inputenc}
\usepackage{amsmath}
\usepackage{amsthm}
\numberwithin{equation}{section}
\usepackage{amsfonts}
\usepackage{mathtools}
\usepackage{mathrsfs}
\usepackage{mathpazo}
\usepackage{amssymb}
\usepackage{graphicx}
\usepackage{ifpdf}
\usepackage[dvipsnames]{xcolor}
\usepackage{multirow}
\usepackage{xcolor}
\usepackage[vcentermath]{youngtab}
\usepackage{tikz}
\usetikzlibrary{arrows.meta,positioning,calc,shapes.geometric}

\makeatletter
\newcommand*\rel@kern[1]{\kern#1\dimexpr\macc@kerna}
\newcommand*\widebar[1]{
  \begingroup
  \def\mathaccent##1##2{
    \rel@kern{0.8}
    \overline{\rel@kern{-0.8}\macc@nucleus\rel@kern{0.2}}
    \rel@kern{-0.2}
  }
  \macc@depth\@ne
  \let\math@bgroup\@empty \let\math@egroup\macc@set@skewchar
  \mathsurround\z@ \frozen@everymath{\mathgroup\macc@group\relax}
  \macc@set@skewchar\relax
  \let\mathaccentV\macc@nested@a
  \macc@nested@a\relax111{#1}
  \endgroup
}
\makeatother
\newcommand{\Cb}{\widebar{C}}
\newcommand{\Qb}{\widebar{Q}}

\usepackage{cite}
\usepackage[bookmarks=true,colorlinks=true,linkcolor=black,citecolor=orange,urlcolor=orange,bookmarksnumbered]{hyperref}

\usepackage[left=2.50cm, right=2cm, top=2cm, bottom=3cm]{geometry}
\begin{document}
\thispagestyle{empty}

\begin{center}

\vspace{-1.0truecm}

{\Large \bf Tensor hierarchy from deformation quantisation}

\vspace{0.5truecm}

{Falk Hassler, David Osten, Alex Swash}

\{\texttt{falk.hassler, david.osten, alex.swash}\}@uwr.edu.pl

\vspace{0.5truecm}

{\em Institute for Theoretical Physics (IFT), \\
University of Wroc\l aw \\
pl. Maxa Borna 9, 50-204 Wroc\l aw, Poland
}

\vspace{0.5truecm}
\end{center}

\begin{abstract}
We show that a formal deformation quantisation of degree-2 differential graded symplectic (QP) manifolds gives rise to the complete classical kinematics and dynamics of NS-NS supergravity, or, equivalently, its duality-covariant formulation as double field theory (DFT). Our construction extends the standard tensor hierarchy, encoding gauge parameters and their redundancies, to a complex encoding the physical fields, field strengths, and Bianchi identities. We present two realisations of such a complex: 1) a differential graded Lie algebra built from the algebra of functions on the QP-manifold; 2) a cochain complex that reproduces the tensor hierarchy representations conjectured in the literature and can be interpreted as a classical BV complex of the linearised theory. The deformation that gives rise to the dilaton comes from a normal-ordered graded Moyal--Weyl star product. This naturally extends the duality structure group from $\mathrm{O}(D,D)$ to $\mathrm{O}(D,D)\times\mathbb{R}^+$. Finally, our framework provides a direct route to constructing the DFT action via local double Lorentz invariance, and offers a transparent algebraic foundation for curvature tensors in generalised Cartan geometry.
\end{abstract}

\tableofcontents

\section{Introduction}

The geometry probed by strings differs fundamentally from the Riemannian geometry experienced by point particles. T-duality, which exchanges momentum and winding modes, obscures the conventional notion of distance and necessitates a generalisation of differential geometry \cite{Giveon:1994fu}. Double field theory (DFT) \cite{Duff:1989tf, Tseytlin:1990nb, Tseytlin:1990va, Siegel:1993xq, Siegel:1993th, Siegel:1993bj, Hull:2004in, Hull:2006va, Hull:2009mi, Hull:2009zb, Hohm:2010jy, Hohm:2010pp, Geissbuhler:2013uka, Aldazabal:2013sca, Berman:2013eva, Hohm:2013bwa} and generalised geometry \cite{Hitchin:2003cxu, Gualtieri:2003dx} provide a powerful framework to make this duality symmetry manifest by extending the tangent bundle to $TM \oplus T^*M$ and promoting its structure group from $\mathrm{GL}(D)$ to $\mathrm{O}(D,D)$. Furthermore, these frameworks combine the metric and Kalb-Ramond B-field into a unified generalised metric, and elevate diffeomorphisms and B-field gauge transformations to generalised diffeomorphisms. However, the algebra of these generalised diffeomorphisms only closes upon imposing the section condition (or strong constraint), which restricts the coordinate dependence of all fields and gauge parameters, effectively halving the dimension of the doubled spacetime and recovering generalised geometry.

A defining feature of these generalised diffeomorphisms is that they are \emph{reducible}: due to the section condition, there exist non-trivial gauge parameters (such as exact one-forms) that generate vanishing gauge transformations \cite{Berman:2012vc}. An algebraic realisation of this reducible structure is the so-called tensor hierarchy $\mathcal{R} = \bigoplus_{n \geq 1} \mathcal{R}_n$. Mathematically, this forms a differential graded Lie algebra (DGLA) concentrated in positive degrees, whose homogeneous subspaces $\mathcal{R}_n$ correspond to certain representations of the underlying structure group, here $\mathrm{O}(D,D)$. This hierarchy cleanly separates the degrees of freedom: $\mathcal{R}_1$ contains the generalised vectors (parameters of generalised diffeomorphisms), while $\mathcal{R}_2$ contains the trivial "gauge-for-gauge" parameters (or ghosts-for-ghosts), and so on. The differential maps higher reducibility parameters to lower ones, for example mapping $\mathcal{R}_2$ to trivial generalised diffeomorphism parameters in $\mathcal{R}_1$.

In general, the natural mathematical language for handling such reducible gauge symmetries is the Batalin-Vilkovisky (BV) formalism \cite{Batalin:1981jr, Batalin:1983ggl}, which is geometrically encoded in the language of $L_\infty$-algebras \cite{Hohm:2017pnh, Jurco:2018sby} and Q-manifolds \cite{Schwarz:1992nx, Alexandrov:1995kv, Stasheff:1997iz}, i.e. graded manifolds endowed with a self-commuting vector field $Q$ of degree 1\footnote{These are also called differential graded (dg) manifolds or $L_\infty$-algebroids in the literature. For the purposes of this article, we will restrict our attention to $\mathbb{N}$-graded manifolds.}. To also encode the $\mathrm{O}(D,D)$ pairing intrinsic to generalised geometry, one must enrich the structure to \emph{symplectic} Q-manifolds, or \emph{QP-manifolds} for short\footnote{Also called symplectic dg-manifolds, NPQ-manifolds or symplectic Lie-$n$ algebroids.}. By a celebrated theorem of Roytenberg \cite{Roytenberg:1999mny, Roytenberg:2002nu, Severa:2017oew}, degree-2 QP-manifolds (QP2) are in one-to-one correspondence with Courant algebroids, the algebraic backbone of generalised geometry \cite{Liu:1995lsa}\footnote{DFT has also been recast in the language of (pre-)QP-manifolds and metric algebroids \cite{Deser:2014mxa, Deser_2018, Deser:2017fko, Deser:2018flj, Heller:2016abk, Carow-Watamura:2018iau}.}. A degree-2 QP-manifold by itself encodes the Courant algebroid structure underlying the generalised tangent bundle and hence the kinematical gauge structure of generalised geometry. It does not yet specify the dynamical NS-NS background: for this one must add a generalised metric, as done here in Section~\ref{sec:dynamics}. In the Alexandrov-Kontsevich-Schwarz-Zaboronsky (AKSZ) construction, the same QP2 data define the topological sector of the string sigma model, with the Hamiltonian encoding the H-flux through a Wess-Zumino coupling \cite{Alexandrov:1995kv, Roytenberg:2006qz}. In this paper we use the QP2-manifold as the algebraic starting point for the NS-NS kinematics and then introduce the additional structures needed for dynamics.

While the standard notion of the tensor hierarchy elegantly resolves the gauge-for-gauge redundancies, this captures only part of the full physical structure.
Hence, one aim and central result of this article is the construction of the following chain complex in the context of QP2-manifolds
\begin{align}
\dots \overset{Q}{\longrightarrow} \hspace{-8pt} \begin{array}{c}
    \mathcal{R}_2 \\ \text{\footnotesize gauge-for-gauge} \end{array} \hspace{-8pt}
 \overset{Q}{\longrightarrow} \hspace{-8pt} \begin{array}{c}
    \mathcal{R}_1 \\ \text{\footnotesize gauge parameters} \end{array} \hspace{-8pt} \overset{Q}{\longrightarrow} \hspace{-5pt} \begin{array}{c}
    \mathcal{R}_0 \\ \text{\footnotesize physical fields} \end{array} \hspace{-8pt} \overset{Q}{\longrightarrow} \hspace{-8pt} \begin{array}{c}
    \mathcal{R}_{-1} \\ \text{\footnotesize field strengths} \end{array} \hspace{-8pt} \overset{Q}{\longrightarrow} \hspace{-8pt} \begin{array}{c}
    \mathcal{R}_{-2} \\ \text{\footnotesize Bianchi identities} \end{array} \hspace{-8pt} \overset{Q}{\longrightarrow} \dots \label{eq:fullchain}
\end{align}
for the duality-covariant formulation of the Neveu-Schwarz--Neveu-Schwarz (NS-NS) sector of supergravity, including the dilaton. Structurally, this complex resembles the \emph{classical BV complex} of the linearised theory: positive degrees encode ghosts and ghosts-for-ghosts, degree zero contains physical variations, and negative degrees contain fields, field strengths and Bianchi identities. In particular, its cohomology captures the non-trivial physical deformations of the theory. Nevertheless, throughout this paper we will refer to this structure simply as the \emph{(extended) tensor hierarchy}.

We will describe two related ways of extending the tensor hierarchy to the non-positive degrees ($\mathcal{R}_{n \leq 0}$), where the physical fields, field strengths, Bianchi identities and so on reside, as predicted in the context of tensor hierarchy algebras and Borcherds superalgebras \cite{Palmkvist:2013vya, Greitz:2013pua, Palmkvist:2015dea, Cederwall:2017fjm, Cederwall:2018aab, Cederwall:2019qnw, Cederwall:2019bai}:
\begin{enumerate}
    \item The full algebra $C^\infty(\mathcal{M})$ of functions on the QP-manifold possesses a strict DGLA structure. The price of keeping this strict algebraic structure is that the negative and zero degree spaces are larger than the conventional tensor hierarchy representations; for example, the degree-zero sector contains diffeomorphisms in addition to the physical fields.

    \item The projected space $\Cb^\infty(\mathcal{M})$, obtained by removing momentum dependence, is not a DGLA, but only a cochain complex. Nevertheless, it will have the natural interpretation as a classical BV complex of the linearised theory. This interpretation will also hold for the generalised Cartan geometry setup in Section \ref{sec:GCG}.
\end{enumerate}

In any case, this usual QP2 framework fails to incorporate the generalised dilaton or its descendants (its field strength and Bianchi identities). While the physical necessity of the dilaton is well-established -- it provides the invariant measure $e^{-2d}$ ensuring the action is invariant under generalised diffeomorphisms -- its geometric origin within graded symplectic geometry remains obscured.
A realisation of such a structure already exists in the generalised geometry and DFT approaches to supergravity -- the frame formulation of DFT using Clifford algebras and Dirac generating operators \cite{Carow-Watamura:2020xij}. In this approach, a Dirac-type operator plays the role of the differential $Q$ in the above complex. It generates all the Courant algebroid structures via derived brackets \cite{Alekseev2001, Severa:2017oew}, and naturally contains the dilaton flux as an ambiguity. Squaring this operator gives us various differential operators, as well as the correct Bianchi identities in DFT. The action is then elegantly obtained by projecting a generalised version of the Lichnerowicz formula \cite{Carow-Watamura:2022ten}. While this Clifford/Dirac perspective is computationally powerful, it operates in a different mathematical regime than graded symplectic geometry.
In this article, we demonstrate that deformation quantisation of the degree-2 QP-manifold provides a precise link between these two pictures\footnote{See \cite{Severa:2017oew, grutzmann2021weyl,keller2015deformation,Severa:2018pag} for mathematical connections between Dirac generating operators and graded geometry, and \cite{Boffo:2019zus, Boffo:2021srg} for classical deformations of QP2-manifolds, without star products.}. By employing a normal-ordered graded Moyal--Weyl star product --- a non-standard choice of deformation --- we deform the exterior algebra of the fermionic fibres into a Clifford algebra. An overview about our approach and the central results is outlined below.
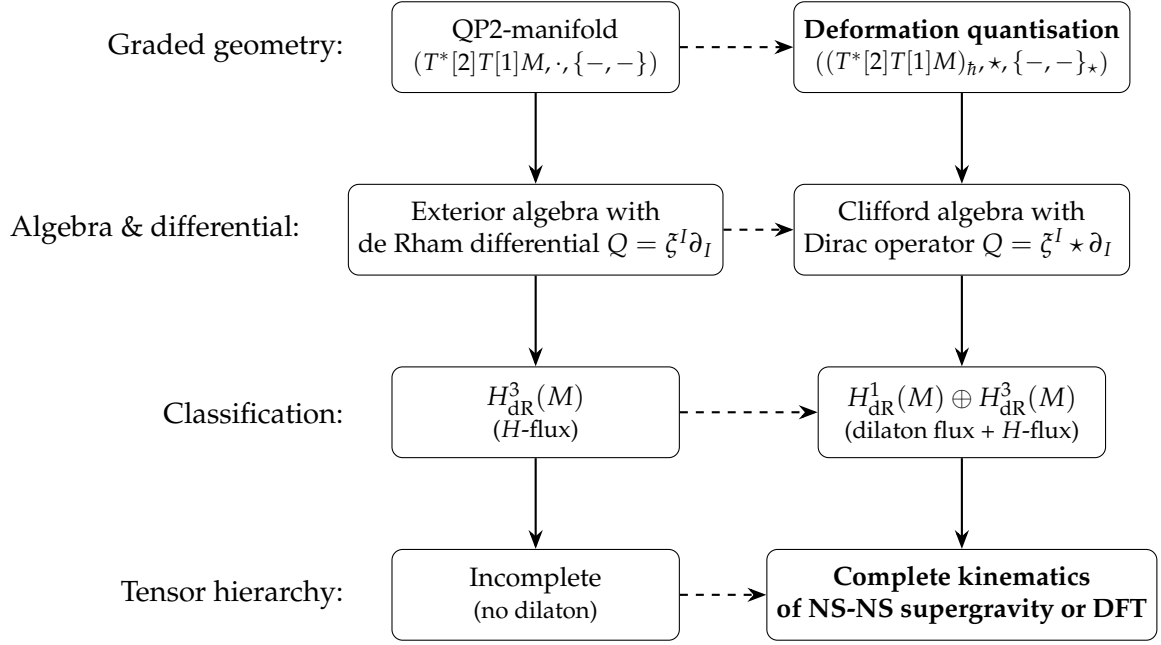
\begin{figure}[htbp]
\centering
\begin{tikzpicture}[
    box/.style={draw, rounded corners, minimum width=3.8cm, minimum height=1.2cm,
                align=center, font=\small},
    arr/.style={-{Stealth[length=2.5mm]}, thick},
    node distance=1.2cm and 1.5cm
]

\node[box] (QP)  {QP2-manifold \\ {\footnotesize $\left( T^*[2]T[1]M , \cdot , \{ - , - \} \right)$}};
\node[box, below=of QP] (dR)  {{Exterior algebra with} \\{de Rham differential $Q = \xi^I \partial_I $}};
\node[box, below=of dR] (Coh1) {$H^3_{\mathrm{dR}}(M)$ \\ {\footnotesize ($H$-flux)}};
\node[box, below=of Coh1] (TH1) {Incomplete \\ {\footnotesize (no dilaton)}};

\node[box, right=of QP] (DQ)  {\textbf{Deformation quantisation} \\ {\footnotesize $\left( (T^*[2]T[1]M)_\hbar , \star , \{ - , - \}_\star \right)$}};
\node[box, below=of DQ] (Cl)  {Clifford algebra with \\ Dirac operator $Q = \xi^I \star \partial_I \  $};
\node[box, below=of Cl] (Coh2) {$H^1_{\mathrm{dR}}(M) \oplus H^3_{\mathrm{dR}}(M)$ \\ {\footnotesize (dilaton flux + $H$-flux)}};
\node[box, below=of Coh2] (TH2) {\textbf{Complete kinematics}\\ \textbf{of NS-NS supergravity or DFT}};

\node[anchor=east] at ([xshift=-0.5cm]QP.west)  {Graded geometry:};
\node[anchor=east] at ([xshift=-0.5cm]dR.west)  {Algebra \& differential:};
\node[anchor=east] at ([xshift=-0.5cm]TH1.west) {{Tensor hierarchy:}};
\node[anchor=east] at ([xshift=-0.5cm]Coh1.west) {Classification:};

\draw[arr, dashed] (QP)  to (DQ);
\draw[arr] (DQ)  to (Cl);
\draw[arr] (QP)  to (dR);
\draw[arr] (Cl)  to (Coh2);
\draw[arr, dashed] (dR)  to (Cl);
\draw[arr, dashed] (TH1) to (TH2);
\draw[arr] (dR)  to (Coh1);
\draw[arr] (Coh1) to (TH1);
\draw[arr] (Coh2) to (TH2);
\draw[arr, dashed] (Coh1)  to (Coh2);

\end{tikzpicture}
\caption{\textit{Schematic depiction of this article's approach: formal deformation quantisation of $T^* [2] T[1] M$ deforms the exterior algebra into a Clifford algebra. This generates the dilaton, completes the tensor hierarchy (yielding the full kinematics of NS-NS supergravity/DFT) and bridges the gap between graded symplectic geometry and the Dirac generating operator approach to DFT.}}
\label{fig:overview}
\end{figure}

The deformation is controlled by an auxiliary, central coordinate $\hbar$ of degree 2. We show that the generalised dilaton is not an independent addition, but arises directly from the generic structure of a degree-2 function in the deformed algebra. Because the star product of two degree-1 coordinates satisfies $\xi^A \star \xi^B = \xi^A \xi^B + \hbar \, \eta^{AB}$, a generic degree-2 generator $\alpha \sim M_{AB} \xi^A \star \xi^B$ inherently contains a trace part proportional to $\hbar$. This trace leads precisely to the dilaton -- in our duality-covariant framework, in the form of the generalised dilaton. Moreover, the deformation quantisation automatically extends the structure group from $\mathrm{O}(D,D)$ to $\mathrm{O}(D,D) \times \mathbb{R}^+$, where the $\mathbb{R}^+$ factor corresponds to the dilaton. The graded commutative product of the QP-manifold is replaced by the star product, and the Poisson bracket is replaced by the star-commutator. The potential Hamiltonian functions $S$ are accordingly deformed: expressing them in terms of the star product introduces an $\hbar$-dependent correction encoding the generalised dilaton flux.  Imposing the deformed master equation determines the generalised fluxes in terms of the generalised frame and yields their Bianchi identities. Crucially, while the master equation largely fixes the one-form flux in terms of the frame's divergence, it leaves a residual gradient ambiguity. We show that resolving this ambiguity naturally requires the introduction of a scalar field -- the generalised dilaton. This reproduces the standard DFT flux relations and completes the tensor hierarchy.

Beyond kinematics, we demonstrate that the deformed framework seamlessly captures the dynamics of NS-NS supergravity. By introducing a generalised metric that breaks the structure group $\mathrm{O}(D,D)$ down to its maximal compact subgroup $\mathrm{O}(D)_L \times \mathrm{O}(D)_R$, we split the Hamiltonian into chiral eigenspaces. We show that requiring local invariance under this subgroup, along with a natural $\mathbb{Z}_2$ symmetry, uniquely fixes the action of DFT, up to an overall normalisation.

This paper is organised as follows. In Section~\ref{sec:QP}, we review the standard QP-manifold approach to Courant algebroids, and demonstrate why its undeformed complex is insufficient to describe the non-positive degree sector of the DFT tensor hierarchy. In Section~\ref{sec:Kinematics}, we introduce the deformation quantisation via a graded star product, derive the emergence of the dilaton, and explicitly construct the full duality-covariant tensor hierarchy. Section~\ref{sec:dynamics} discusses the dynamics, showing how the introduction of a generalised metric and local invariance under its structure group uniquely determine the DFT action. In Section~\ref{sec:GCG}, we apply the deformed QP-manifold structure to the covariant formulation of generalised Cartan geometry. Finally, we conclude in Section~\ref{sec:Outlook} with a brief outlook on applications and open directions.

\section{Tensor hierarchy from QP-manifolds} \label{sec:QP}
A QP-manifold of degree $n$ is a (non-negatively) graded manifold $\mathcal{M}$ equipped with a symplectic structure (the P-structure) $\omega$ of degree $n$ and a homological vector field $Q$ of degree one (the Q-structure), which satisfy $L_Q \omega = 0$ and $L_Q Q = [Q,Q] = 0$. Due to the first condition, $Q$ is compatible with the graded Poisson bracket that arises from $\omega$. The second condition guarantees that $Q$ is a homological vector field because it is nilpotent, $Q^2 = \frac{1}{2} L_Q Q  = 0$. Smooth functions $C^\infty (\mathcal{M})$ on $\mathcal{M}$ form a direct sum of functions of homogeneous degree, and form a Poisson algebra. The graded Poisson bracket of these functions is induced from the symplectic structure via 
\begin{equation}
    \{ f, g \} = (-1)^{|f|+n+1} \iota_{X_f} \iota_{X_g} \omega,
\end{equation}
for $f,g \in C^\infty (\mathcal{M})$, where $X_f$ is the Hamiltonian vector field defined by $\iota_{X_f} \omega = - \mathrm{d} f$. 
It turns out that for any QP-manifold, one can find a Hamiltonian function $S$ of degree $n+1$ associated to $Q$ such that
\begin{equation}
    Q f = \{ S, f \} 
\end{equation}
for $f \in C^\infty (\mathcal{M})$. The nilpotency of $Q$ now translates to the fact that the Hamiltonian function $S$ satisfies the so-called \emph{classical master equation},
\begin{equation}
    \{ S, S \} = 0.
\label{eq:CME}
\end{equation}
For more definitions and properties of Q(P)-manifolds, we refer to the reviews \cite{Cattaneo:2010re, Ikeda:2012pv, 10.1007/978-3-031-89857-0_10, cueca2026lecturenotessymplecticgeometry}.

One can also twist QP-manifolds of degree $n$ by a canonical transformation. Let us define the exponential adjoint action on a smooth function $f$,
\begin{equation}\label{eq:canonical}
    e^{\delta_\alpha} f = f + \{ f, \alpha \} + \frac12 \{ \{ f, \alpha \} , \alpha \} + \dots ,
\end{equation}
where $\alpha \in C^\infty (\mathcal{M})$ is of degree $n$ and $\delta_\alpha = \{ \cdot , \alpha \}$. Moreover, it is a canonical transformation, so it satisfies
\begin{equation}\label{eq:canonicaltr}
    \{ e^{\delta_\alpha} f, e^{\delta_\alpha} g \} = e^{\delta_\alpha} \{ f,g \} ,
\end{equation}
for smooth functions $f,g$. Consequently, if a Hamiltonian function $S$ satisfies the classical master equation, then the twisted Hamiltonian $e^{\delta_\alpha} S$ does as well.

\paragraph{Courant algebroids.} A celebrated result \cite{Roytenberg:1999mny,Severa:2017oew} establishes that QP-manifolds of degree 2 are in one-to-one correspondence with Courant algebroids, which characterise generalised geometry. Given a $D$-dimensional manifold $M$, a Courant algebroid is defined as a vector bundle $E \rightarrow M$ equipped with three operations: a non-degenerate inner product on the fibres $\langle \cdot, \cdot \rangle$, an anchor map to the tangent bundle $\rho : E \rightarrow TM$, and a so-called Dorfman bracket on its sections $[\cdot, \cdot]$. These must satisfy the following compatibility conditions \cite{Liu:1995lsa},
\begin{subequations}
\label{eq:CA}
\begin{align}
    [e, f e' ] &= f [e, e'] + ( \rho(e) f )  \, e' , \\
    [e, [e', e'']] &= [[e, e'], e''] + [ e', [e, e'']] , \\
    \rho (e) \langle e', e'' \rangle &= \langle [e, e'], e'' \rangle + \langle e', [ e, e'' ] \rangle , \\
    [e, e'] + [e', e] &= \rho^* \mathrm{d} \langle e, e' \rangle, 
\end{align}
\end{subequations}
where $e, e', e'' \in \Gamma (E)$ and $f \in C^\infty (M)$. The prototypical example is the \emph{standard Courant algebroid}, defined on the generalised tangent bundle $E = TM \oplus T^* M$ equipped with its canonical symmetric pairing, the natural projection to $TM$ as the anchor map, and the standard Dorfman bracket. 

The correspondence to QP-manifolds of degree 2 proceeds as follows: consider a vector bundle $A \rightarrow M$ and the graded manifold $\mathcal{M} = T^* [2] A[1]$. The degree-1 coordinates arise from the fibres of $A$ and their conjugate momenta in $A^*$, spanning a total vector bundle $E = A \oplus A^*$. We denote the local coordinates of $\mathcal{M}$ as $(x^i, \xi^A, p_i)$ of respective degrees $(0,1,2)$, where $\xi^A$ are the coordinates on $E[1]$, and $x^i$ are the coordinates on $M$.  The canonical symplectic structure naturally equips the fibres of $E$ with a non-degenerate pairing $\langle \xi^A , \xi^B \rangle = \eta^{AB}$, yielding the symplectic form
\begin{equation}
    \omega = \mathrm{d} x^i \wedge \mathrm{d} p_i + \frac12 \eta_{AB} \, \mathrm{d} \xi^A \wedge \mathrm{d} \xi^B \, .
\end{equation}
The most general degree-3 Hamiltonian on this space takes the form
\begin{equation}
    S = \rho_A^i (x) p_i \xi^A+ \frac{1}{3!} F_{ABC}(x) \xi^A \xi^B \xi^C, \label{eq:generalform}
\end{equation}
where $\rho_A^i, F_{ABC} \in C^\infty (M)$. Requiring this Hamiltonian to solve the classical master equation \eqref{eq:CME} perfectly reproduces the defining axioms \eqref{eq:CA} of a Courant algebroid, provided we identify $\rho_A^i$ as the anchor map (viewed as a section $\rho \in \Gamma ( E^* \otimes TM )$) and $F \in \Gamma (\wedge^3 E^*)$ as the structure functions of the Dorfman bracket in a suitable basis, the so-called 'generalised fluxes'
\begin{equation}
    F_{ABC} = E_{[\underline{A}}{}^M \partial_M E_{\underline{B}}^N E_{N\underline{C}]}, \label{eq:GeneralisedFluxes}
\end{equation}
where $E_A{}^M \in \mathrm{O}(D,D)$ is a generalised frame of the generalised tangent bundle, and $\rho_A^i = E_A{}^i$. The defining objects of the Courant algebroid are then recovered via derived brackets \cite{kosmann2004derived}:
\begin{align}
    [ e, e' ] &= - \{ \{ e, S \}, e' \}, \nonumber \\
    \rho(e) f &= - \{ \{ e, S \}, f \}, \label{eq:DerivedBrackets}\\
    \langle e, e' \rangle &= \{ e, e' \}, \nonumber
\end{align}
where $f \in C^\infty (M) $ (degree $0$ on $\mathcal{M}$) and $e, e'$ are smooth functions of degree $1$ on $\mathcal{M}$, identified with sections of $E$.

\subsection{\texorpdfstring{The complex of functions on $\mathcal{M}$ as a differential graded Lie algebra}{The complex of functions on M}} \label{sec:DGLAextension}

The Poisson algebra $C^\infty (\mathcal{M})$ of smooth functions on the QP-manifold $\mathcal{M}$ possesses a natural differential graded Lie algebra (DGLA) structure \cite{Ritter:2015ffa, Arvanitakis:2018cyo}. Let us first recall that a DGLA $X$ is a graded vector space equipped with a graded Lie bracket $[\cdot, \cdot]$ of degree $0$, and a differential $D$ of degree $-1$ satisfying the graded Leibniz rule: 
\begin{equation}
    D[a,b] = [Da, b] + (-1)^{\text{deg} \: a} [a, Db]\,. \label{eq:derivation}
\end{equation}
The space of homogeneous functions of degree $m$ on the QP-manifold, denoted by $C^\infty_m (\mathcal{M})$, is then identified with the subspace $X_{n-m}$ of degree $n-m$ in the DGLA, where $n$ is the degree of $\mathcal{M}$. Schematically, a reflection in degree gives the correspondence:
\begin{equation}
    \begin{aligned}
        \{\, \cdot\,,\, \cdot \,\} \quad &\rightarrow \quad [\, \cdot\,,\, \cdot \,] \\
        Q  \, \quad &\rightarrow \quad D \,.
    \end{aligned}
\end{equation}
\paragraph{The tensor hierarchy.} A physical incarnation of a DGLA appears in supergravity theories and extended geometries as the tensor hierarchy of a duality group $\mathcal{G}$. Here, the subspaces $X_n$ correspond to representations $\mathcal{R}_n$ of $\mathcal{G}$ \cite{Palmkvist:2013vya, Greitz:2013pua, Lavau:2019oja, Bonezzi:2019bek}, or their associated vector bundles\footnote{When there is no risk of confusion, we will use the notation $\mathcal{R}_n$ for both the representation and the bundle.}. The vector space $\mathcal{R}_1$ typically corresponds to sections of a vector bundle $E$, on which we can define a Dorfman bracket via a derived bracket. In this paper, we focus on the duality groups $\mathcal{G} = \mathrm{O}(D,D)$ and\footnote{The $\mathbb{R}^+$-factor accounts for the fact that we can attribute a density weight as well.} $\mathcal{G} = \mathrm{O}(D,D) \times \mathbb{R}^+$. The corresponding representations suggested in the literature are:
\begin{equation}
    \mathcal{R}_2 = \mathbf{1}, \qquad \mathcal{R}_1 = \mathbf{2D}, \qquad \mathcal{R}_0 = \mathbf{adj} , \quad \dots
\end{equation}
where $\mathbf{adj}$ is the adjoint representation of the Lie algebra of $\mathcal{G}$.

Motivated by the structure \eqref{eq:fullchain}, we aim to establish the correspondence
\begin{equation}
    \mathcal{R}_n \cong \Cb^\infty_{2-n}(\mathcal{M}), \qquad \forall \, n \leq 2
\end{equation}
for some class of functions on the QP-manifold defined below. To achieve this, we must extend the hierarchy in the non-positive sector ($\mathcal{R}_{n\leq 0}$) to capture the physical degrees of freedom, their field strengths, and Bianchi identities. The physical interpretation of these spaces is well established, notably within the context of tensor hierarchy algebras \cite{Cederwall:2021xqi, Cederwall:2023xbj, Cederwall:2025iyq}: $\mathcal{R}_0$ houses the duality group's Lie algebra (the physical fields): $\mathcal{R}_{-1}$ houses the embedding tensors, identified with the generalised fluxes in the frame formulation of DFT; $\mathcal{R}_{-2}$ houses their corresponding Bianchi identities, and so on. While the algebraic structure of these negative degrees is understood in certain settings, an explicit realisation has been missing.

We will see that extending the tensor hierarchy to the non-positive degree sector $\mathcal{R}_{n\leq 0}$ is quite subtle, even for simple duality groups like $\mathrm{O}(D,D)$ or $\mathrm{O}(D,D)\times \mathbb{R}^+$. It is possible to maintain the DGLA structure of the positive-degree sector, but for this one needs to enhance the physically motivated choices for the non-positive representations $\mathcal{R}_{n\le0}$. Although the Lie algebra structure on $C^\infty(\mathcal{M})$ carries over easily to $\Cb^\infty(\mathcal{M})$, this is not true for the differential. Motivated by the underlying physics, we will therefore take a different, homological, point of view to overcome this problem.

\paragraph{The realisation on the QP-manifold.} For the QP2-manifold $\mathcal{M} = T^*[2]T[1]M$ modelling the standard Courant algebroid on $E = TM \oplus T^* M$, we have the following isomorphisms: 
\begin{equation}
    C^\infty_1 (\mathcal{M}) \cong \Gamma (E) = \mathcal{R}_1 \quad \text{and} \quad C^\infty_0 (\mathcal{M}) \cong C^\infty (M) = \mathcal{R}_2.
\end{equation}
This corresponds to the \textit{positive-degree} sector of the tensor hierarchy DGLA of generalised geometry/double field theory \cite{Arvanitakis:2018cyo}, to which one can canonically associate an $L_\infty$-algebra \cite{roytenberg1998courant, voronov2005higher, fiorenza2007structures, getzler2010higher}.

It turns out that the space $C^\infty_2 (\mathcal{M})$, corresponding to canonical transformations of the graded Poisson bracket, does not match with $\mathcal{R}_0 = \mathfrak{o}(D,D) \oplus \mathbb{R}$. This mismatch arises for two reasons:
\begin{itemize}
    \item The space $C^\infty_2 (\mathcal{M})$ contains functions of the degree-2 coordinates $p$ that have no corresponding part in $\mathcal{R}_0$. These will be removed by projecting to an appropriate subspace $\Cb^\infty(\mathcal{M}) \subset C^\infty (\mathcal{M})$, which we define below in Section \ref{sec:reduced}. Nevertheless, we will see that only the complete complex $C^\infty_2 (\mathcal{M})$ including functions depending on the $p$-coordinates will give rise to a DGLA.
    \item There is no coordinate or degree of freedom in the construction that would account for the singlet part of $\mathcal{R}_0$, which corresponds to the dilaton. Resolving this second issue will require a formal deformation quantisation, which we introduce in Section \ref{sec:Kinematics}.
\end{itemize}
Indeed, the most general degree-2 function is expressed in local coordinates as
\begin{equation}
    \alpha = v^i (x) p_i + \frac12 M_{IJ} (x) \xi^I \xi^J. 
\end{equation}
We observe that the first term, being a section of $TM$, generates infinitesimal diffeomorphisms of $M$, while the second, due to $\wedge^2 E \cong \mathfrak{o}(D,D)$, is responsible for infinitesimal $\mathrm{O}(D,D)$ transformations. This is seen explicitly by looking at the adjoint action of $\alpha$ on the coordinates,
\begin{align}
    \delta_\alpha x^i &= \{ x^i, \alpha \} = v^i , \nonumber \\
    \delta_\alpha \xi^I &= \{ \xi^I, \alpha \} = M_J\mathstrut^I \xi^J , \label{eq:TransformationExt} \\
    \delta_\alpha p_i &= \{ p_i, \alpha \} = - \partial_i v^j p_j - \frac12 \partial_i M_{JK} \xi^J \xi^K, \nonumber
\end{align}
which integrate to 
\begin{align}
    x'^i &= x'^i (x) , \nonumber \\
    \xi'^I &= E_J\mathstrut^I (x) \xi^J , \label{eq:pTransformationUndeformed} \\
    p'_i &= \frac{\partial x^j}{\partial x'^i} p_j + \frac12 \partial_i E_J\mathstrut^L E_{KL} \xi^J \xi^K = \frac{\partial x^j}{\partial x'^i} p_j + \frac{1}{2} \Omega_{iJK} \xi^J \xi^K ,  \nonumber
\end{align}
describing reparametrisations $x^\prime(x)$ and the introduction of a generalised frame ${E_J}^I \in \mathrm{O}(D,D)$ on the fibres of the vector bundle $E$. We also note that the transformation of momenta $p$ includes a non-linear term involving the Weitzenb\"ock connection $\Omega_{IJ}{}^K = \partial_I E_J\mathstrut^L E_{L}\mathstrut^K$, due to the introduction of a generalised frame on $E$, in addition to the standard transformation of $p$ as a covector \cite{Roytenberg:2002nu}. To summarise: we saw that the degree-2 functions on $\mathcal{M}$ now include elements of the form $v^i(x) p_i$ (generating diffeomorphisms) alongside the $\mathfrak{o}(D,D)$-valued functions. The corresponding degree-0 representation $\widetilde{\mathcal{R}}_0$ is therefore not simply $\mathfrak{o}(D,D)$, but rather the \emph{Atiyah algebra}:
\begin{equation}
    \widetilde{\mathcal{R}}_0 = \mathfrak{X}(M) \ltimes \mathfrak{o}(D,D),
\end{equation}
which is the Lie algebra of infinitesimal automorphisms of the Courant algebroid \cite{Roytenberg:2002nu}. The price for recovering the DGLA structure is the loss of pure $\mathrm{O}(D,D)$-covariance: the degree-0 space includes diffeomorphisms on the base manifold, rather than physical $\mathrm{O}(D,D)$ representations. In this way, our formalism reconciles naturally with the claim in \cite{Bonezzi:2019bek} that there exists a DGLA in which all $\mathcal{R}_n$ are representations of the degree-0 space. Because the tensor hierarchy bundles $\mathcal{R}_n$ are constructed over the base manifold $M$, they transform naturally under both $\mathfrak{o}(D,D)$ and local diffeomorphisms $\mathfrak{X}(M)$. Thus, the full Atiyah algebra $\widetilde{\mathcal{R}}_0$ acts canonically on all $\mathcal{R}_p$, as demonstrated explicitly in \eqref{eq:TransformationExt}.

Let us also note, that one can also straightforwardly compute the associated homological vector field 
\begin{align*}
    Q &= - \xi^A E_{A}{}^I \frac{\partial}{\partial x^I} + \left(\eta^{AB} E_B{}^M p_M + \frac{1}{2} F^A{}_{BC} \xi^B \xi^C \right) \frac{\partial}{\partial \xi^A} \\
    &{} \quad + \left( \partial_M E_A{}^N p_N \xi^A + \frac{1}{3!} \partial_M F_{ABC} \xi^A \xi^B \xi^C \right) \frac{\partial}{\partial p_M}
\end{align*}
to the standard Hamiltonian \eqref{eq:generalform} expressed in a general generalised frame $E_A{}^M$. By construction, this vector field is homological. We emphasise this, as for the complex proposed in the next section, a choice of $Q$ for an arbitrary generalised frame is not possible.

\paragraph{The standard complex and its cohomology.} As we saw, $C^\infty_2(\mathcal{M})$ does not directly correspond to the expected $\mathcal{R}_0$ in generalised geometry, due to the dependence on the momentum coordinates $p$. Indeed, we have the following isomorphism (the right hand side is called the \emph{Rothstein algebra} \cite{keller2015deformation}),
\begin{equation}
    C^\infty_n (\mathcal{M}) \cong \bigoplus_{2i+j = n} \Gamma \big( \text{Sym}^i (TM) \otimes \wedge^j E^* \big),
\end{equation}
where the functions depending on the $p$ coordinates are responsible for the first factor. The pair  $(C^\infty (\mathcal{M}), Q)$ forms a cochain complex called the \emph{standard complex} of a Courant algebroid. One can then study its cohomology $H^\bullet ( C^\infty (\mathcal{M}), Q)$, which was proven to be isomorphic to the de Rham cohomology of $M$ \cite{Roytenberg:2002nu}. For other discussions of Courant algebroid cohomology, see \cite{Lyakhovich:2004kr, Lyakhovich:2009qq, GinotGrutzmann2009, keller2015deformation, CuecaMehta2021}.

To understand this result, consider the splitting $\xi^I = (\xi^i, \xi_i)$ and the differential $Q_0$ associated to the flat Hamiltonian $S_0 = \xi^i p_i$. Under this differential, we observe that $Q_0 (\xi_i) = p_i$ and $Q_0 (p_i) = 0$. This implies that the fibre coordinates $(\xi_i, p_i)$ form a \emph{contractible pair}, and thus drop out of the Q-cohomology entirely \cite{Barnich:2000zw}. To see this, we introduce the contracting homotopy operator $\rho = \xi_i \frac{\partial}{\partial p_i}$. Using this, we define the counting operator $N$ via the anti-commutator
\begin{equation}
    N = Q_0 \rho + \rho Q_0 \, .
\end{equation}
By acting on the coordinates, one finds that $N(x^i) = N(\xi^i) = 0$, while $N(\xi_i) = \xi_i$ and $N(p_i) = p_i$. Therefore, the operator $N$ strictly counts the total number of the contractible variables $(\xi_i, p_i)$ in any given polynomial. Now, consider a general $Q_0$-closed function $F_k$ that contains exactly $k \geq 1$ of these variables, i.e. $N(F_k) = k \, F_k$. Applying the definition above, we also have
\begin{equation}
    k \, F_k = ( Q_0 \rho + \rho Q_0 ) F_k = Q_0 (\rho F_k) + \rho ( Q_0 F_k) \, .
\end{equation}
Because we assumed that $F_k$ is closed, this proves that $F_k$ is in fact $Q_0$-exact:
\begin{equation}
    F_k = Q_0 \left( \frac1k \rho F_k \right) \, .
\end{equation}
The cohomology of the complex is therefore determined by the remaining coordinates $(x^i, \xi^i)$, parametrising the manifold $T[1]M$. We then use the following isomorphism,
\begin{equation}
    C^\infty (T[1]M) \cong \Omega^\bullet (M) \, ,
\end{equation}
under which $Q_0$ reduces to the exterior derivative. This proves that $H^\bullet ( C^\infty (\mathcal{M}), Q_0) \cong H^\bullet_{\text{dR}} (M)$, and demonstrates that the additional structure in the QP-manifold is cohomologically trivial, in particular the momentum coordinates $p_i$. This match of cohomologies extends from $Q_0$ to any other $Q$ obtained by canonical transformations \eqref{eq:canonicaltr}, since these are quasi-isomorphisms of the cochain complex. The original proof is given rigorously in \cite{Roytenberg:2002nu}.

\subsection{\texorpdfstring{A reduced complex of functions on $\mathcal{M}$ as a cochain complex}{A reduced complex of functions on M as a cochain complex}}
\label{sec:reduced}
Taking into account the fact that the momenta $p_i$ cannot encode non-trivial cohomology data, which is relevant for the physical observables, we restrict our attention to the space of functions $f$ on $\mathcal{M}$ that do not depend on $p_i$,
\begin{equation}
    \Cb^\infty (\mathcal{M}) = \left\{ f \in C^\infty (\mathcal{M}) \; \big| \; \frac{\partial f}{\partial p_i} = 0 \right\}.
\end{equation}
This restriction removes the $p$-dependent functions that cause the overcounting of $C^\infty(\mathcal{M})$ relative to the $\mathrm{O}(D,D)$ representations typically used in the tensor hierarchy of NS-NS supergravity. Strictly speaking, one could also restrict to functions that do not depend on the cotangent coordinates $\xi_i$, keeping only $\xi^i$ -- this would yield the standard de Rham complex on $M$\cite{Roytenberg:2002nu}. However, we choose to retain the full doubled coordinate $\xi^I = (\xi^i, \xi_i)$ in order to maintain the manifest $\mathrm{O}(D,D)$ structure of the generalised tangent bundle $E = TM \oplus T^*M$. 

Geometrically, $\Cb^\infty(\mathcal{M})$ consists of functions of $(x^i, \xi^I)$, which are precisely functions on the shifted generalised tangent bundle $\Cb^\infty(\mathcal{M}) \cong C^\infty (E [1]) \cong \Gamma (\wedge^\bullet E^*) \, $. This can be viewed as an 'extended' de Rham complex or exterior algebra of the \emph{generalised} tangent bundle.
\begin{table}[ht]
\centering
\begin{align*}
{\renewcommand{\arraystretch}{1.5}
\begin{array}{c|ccc|c|c}
\text{Functions} & \multicolumn{3}{|c|}{\text{Representation}} & \text{Local expression} & \text{Physical meaning} \\ \hline \hline 
    \Cb^\infty_0(\mathcal{M}) & \mathcal{R}_2 &=& \bullet &  \lambda(x) & \text{gauge for gauge transformations} \\ \hline
    \Cb^\infty_1(\mathcal{M}) & \mathcal{R}_1 &=& \yng(1) &  V_I(x) \xi^I & \text{gauge transformations} \\ \hline
    \Cb^\infty_2(\mathcal{M}) & \mathcal{R}_0 &=& \raisebox{-15pt}{\rule{0pt}{35pt}}\yng(1,1)  & \frac12 M_{IJ}(x) \xi^I \xi^J  & \text{gauge fields (generalised frame)} \\ \hline
    \Cb^\infty_3(\mathcal{M}) & \mathcal{R}_{-1} &=& \raisebox{-20pt}{\rule{0pt}{45pt}}\yng(1,1,1) &  \frac{1}{3!} F_{IJK}(x) \xi^I \xi^J \xi^K& \text{field strengths} \\ \hline
    \Cb^\infty_4(\mathcal{M}) & \mathcal{R}_{-2} &=& \raisebox{-25pt}{\rule{0pt}{55pt}}\yng(1,1,1,1) &  \frac{1}{4!} Z_{IJKL}(x) \xi^I \xi^J \xi^K \xi^L & \text{Bianchi identities \& sources} \\ \hline
    \vdots & \multicolumn{3}{c|}{\vdots} & \vdots & \vdots
\end{array}}
\end{align*}
    \caption{Tensor hierarchy associated to the Courant algebroid/QP-manifold $T^* [2]T[1]M$. The Young tableaux correspond to representations of $\mathrm{O}(D,D)$.}
    \label{tab:THwoDilaton}
\end{table}
The space $\Cb^\infty (\mathcal{M})$, equipped with the following restricted differential
\begin{equation}
    \Qb = Q_0 |_{p=0} = - \xi^i \frac{\partial}{\partial x^i} \, ,
\end{equation}
also forms a cochain complex. Moreover, the complex $\Cb^\infty (\mathcal{M})$ is closed under the Poisson bracket. In total, we will define
\begin{equation}
    (\Cb^\infty(\mathcal{M}) , \Qb , \{ \cdot , \cdot \} )
\end{equation}
as another completion of the tensor hierarchy to negative degrees. Let us note the several issues that this definition of the tensor hierarchy has:
\begin{itemize}
    \item Although this complex is both a Poisson algebra and a cochain complex, it is in general, \textit{not} a DGLA any more -- the reason being that $\Qb$ is not a derivation of the Poisson bracket.

    \item The homological vector field $\Qb$ cannot be expressed in an arbitrary generalised frame. The natural candidate for such an object given by
    \begin{equation}
        \Qb = - \xi^A E_A{}^M \partial_M + \frac{1}{2} F^A{}_{BC} \xi^B \xi^C \frac{\partial}{\partial \xi^A}
    \end{equation}
    fails to be homological if the generalised fluxes $F^A{}_{BC}$ are not constant in general.
    \item $\Qb$ is generically not a Hamiltonian vector field.
\end{itemize}
All of these properties can be restored if one of the two additional conditions are satisfied:
    \begin{enumerate}
        \item One further restricts the space of functions $\Cb^\infty(\mathcal{M})$ by dropping all dependence on the base coordinates $x^i$. Equivalently, one restricts $\mathcal{M}$ to the submanifold $\mathfrak{f}[1]$, where $\mathfrak{f}$ is a $2D$-dimensional quadratic Lie algebra, embedded into $\mathfrak{o}(D,D)$. Conceptually, this acts as a symplectic reduction of the QP-manifold, reproducing the tensor hierarchy algebra \cite{Palmkvist:2013vya, Greitz:2013pua}. The Hamiltonian restricts to $\widebar{S} = \frac{1}{3!} F_{ABC} \xi^A \xi^B \xi^C \,$, where the master equation implies the Jacobi identity for the structure constants $F_{ABC}$. The gauge-inequivalent choices of $\widebar{S}$ are strictly classified by the Chevalley-Eilenberg Lie algebra cohomology $H^3_{\text{CE}}(\mathfrak{f}, \mathbb{R})$. The resulting space of functions $C^\infty (\mathfrak{f}[1])$ is isomorphic to the Chevalley-Eilenberg complex of $\mathfrak{f}$, and the graded Poisson bracket upgrades this complex to a DGLA.

        \item  \label{item:minimal} Alternatively, one can drop the $\xi_i$ dependence of the functions, implying that we restrict to the submanifold $T[1]M$. The remaining coordinates $(x^i,\,\xi^i)$ govern the standard de Rham complex, and the Poisson bracket trivialises. In this case, we are not able to express $\Qb$ in a generic generalised frame, or equivalently twist by an arbitrary $\mathrm{O}(D,D)$ element anymore.\footnote{$\Qb$ possesses only the trivial geometric twist by an ordinary frame ${e_a}^m$ on $TM$: $\Qb = - \xi^a e_a{}^m \partial_m + \frac{1}{2} f^a{}_{bc} \xi^b \xi^c \frac{\partial}{\partial \xi^a}.$
        Another compatible twist of a generalised frame is the twist by a $B$-field. For this let us note, that us in the case of $C^\infty(\mathcal{M})$, the $H$-flux does not affect the $\Qb$ but only the Hamiltonian, if acting on $T[1]M$.}
\end{enumerate}
Naturally, if one is strictly interested in the cohomology of this complex, it collapses, modulo spectator coordinates, to the standard de Rham cohomology $H^\bullet_{\text{dR}}(M)$, as demonstrated by Roytenberg \cite{Roytenberg:2002nu}. In this sense, the complex $\Cb^\infty(\mathcal{M})$ is the intermediate step between the DGLA $C^\infty(\mathcal{M})$ and the de Rham complex on $M$.

The main reason for considering this reduced complex $(\Cb^\infty(\mathcal{M}), \Qb)$ is that it consists only of the pure $\mathrm{O}(D,D)$ representations and yields exactly what is typically identified as the tensor hierarchy in the physical literature (see Table~\ref{tab:THwoDilaton}). A physical motivation to restore the DGLA properties is that, in a linearised theory\footnote{In the sense that elements of the tensor hierarchy only appear linearly.}, the Poisson brackets are rendered trivial. In that case the generalised fluxes \eqref{eq:GeneralisedFluxes} trivialise. Consequently, the complex associated to Table~\ref{tab:THwoDilaton} should be associated to the physical objects of the \textit{linearised theory}. Moreover, the complex presented there is the \textit{classical BV complex of the linearised theory.} The same will be true for generalised Cartan geometry, considered in Section \ref{sec:GCG}.

To summarise this section, the unrestricted complex $C^\infty(\mathcal{M})$ is a completion to a DGLA -- but its representations require an enhancement in comparison to the standard tensor hierarchy, whereas $\Cb^\infty(\mathcal{M})$ is not a DGLA but reproduces the tensor hierarchy for negative degrees known in the literature. Furthermore, it gives rise to a classical BV complex, at least of the linearised theory. The same issue will arise when considering the deformation quantisation of this setup in the next section. 

\paragraph{Absence of the dilaton.} Nevertheless, the structure in Table \ref{tab:THwoDilaton} is \textit{not} sufficient to describe the complete kinematics and dynamics of the NS-NS sector of supergravity in a duality-covariant formulation. Whereas for $\mathcal{R}_2 = C^\infty_0 (\mathcal{M}) = \Cb^\infty_0 (\mathcal{M})$ and $\mathcal{R}_1 = C^\infty_1 (\mathcal{M}) = \Cb^\infty_1 (\mathcal{M})$ one reproduces the standard DFT tensor hierarchy \cite{Hohm:2013nja}, this pattern does not continue for $\mathcal{R}_{n \leq 0}$ with the present structure. If we want the full space $\Cb^\infty (\mathcal{M}) = \bigoplus_n \Cb^\infty_n (\mathcal{M})$ to correspond to the DFT tensor hierarchy, then $\Cb^\infty_2 (\mathcal{M})$ should be $\mathcal{R}_0 \cong \mathfrak{o}(D,D) \oplus \mathbb{R}$, corresponding to the physical fields (generalised frame ${E_A}^I$ and dilaton $d$). Moreover, $\Cb^\infty_3 (\mathcal{M})$ should host their field strengths $F_{ABC}$ and $F_A$, furnishing the representation $\mathcal{R}_{-1}$. The representation $\mathcal{R}_{-2}$, corresponding to the space of Bianchi identities for these field strengths, should be identified with $ \Cb^\infty_4(\mathcal{M}) $, and so on. In the following section, we will see how to incorporate the (generalised) dilaton by deforming the Poisson bracket, and how this will lead to the full DFT tensor hierarchy.

\section{Kinematics} \label{sec:Kinematics}
As discussed above, the complexes $\big( \Cb^\infty, \Qb \big) $ and $\big( C^\infty, Q \big) $ associated with the QP2-manifold $\mathcal{M} = T^*[2] T[1] M$ provide completions of the DFT tensor hierarchy down to negative degrees. However, as noted, they fail to capture the generalised dilaton. In this section, we demonstrate that the dilaton and its descendants (flux, Bianchi identities and so on) emerge naturally when we consider a formal deformation quantisation of $\mathcal{M}$.

\subsection{Deformation of Poisson bracket via star product}
In the following, we consider the QP-manifold of degree two $\mathcal{M} = T^* [2] T[1]M$ describing standard Courant algebroids.
The classical graded Poisson bracket of two smooth homogeneous functions $f,g$ of respective degrees $|f|, |g|$ is:
\begin{equation}
    \{ f, g \} = \frac{\partial f}{\partial x^i} \frac{\partial g}{\partial p_i} - (-1)^{|f| |g|} \frac{\partial g}{\partial x^i} \frac{\partial f}{\partial p_i} - (-1)^{|f|} \eta^{IJ} \frac{\partial f}{\partial \xi^I} \frac{\partial g}{\partial \xi^J}.
\label{eq:ClassicalPoisson}
\end{equation}
In the following, we will denote the right derivative with respect to the degree-1 coordinates by $f \frac{\overset{\leftarrow}{\partial}}{\partial \xi^I} = -(-1)^{|f|} \frac{\partial f}{\partial \xi^I}$.
Classically, these coordinates $\xi$ anti-commute, generating an exterior algebra or Grassmann algebra, and have a Poisson bracket giving rise to the Courant algebroid pairing, $\{ \xi^I, \xi^J \} = \eta^{IJ}$. We can unify these structures by deforming the exterior algebra into the Clifford algebra via deformation quantisation \cite{BAYEN197861, 1983LMaPh...7..487D, Fedosov:1994zz, Kontsevich:1997vb, Bordemann:1999ca, Tyutin:2001iz, Hirshfeld:2002ki}. We implement this via a star product (which will be defined explicitly below), 
\begin{equation}
    \xi^I \star \xi^J = \xi^I \xi^J + \hbar \eta^{IJ},
\end{equation}
where $\hbar$ is an auxiliary coordinate of degree two, introduced to preserve the grading. The deformed Poisson bracket is defined as the graded star-commutator, normalised by $\frac{1}{2\hbar}$ to maintain degree $-2$,
\begin{equation}
    \{ f, g \}_\star = \frac{1}{2\hbar} ( f \star g - (-1)^{|f| |g|} g \star f ).
\end{equation}
The fibre metric is recovered via the star-commutator of two degree-1 coordinates,
\begin{equation}
    \{ \xi^I, \xi^J \}_\star = \eta^{IJ}. \label{eq:deg1stardefPB}
\end{equation}
The star product we will consider is the (normal-ordered) graded Moyal--Weyl product of functions $f,g \in C^\infty (\mathcal{M})$,
\begin{align}
    f \star g &= f \exp  \hbar \left(2  \overset{\longleftarrow}{\frac{\partial}{\partial x^I}} \overset{\longrightarrow}{\frac{\partial}{\partial p_I}} + \overset{\longleftarrow}{\frac{\partial}{\partial \xi^I}} \eta^{IJ} \overset{\longrightarrow}{\frac{\partial}{\partial \xi^J}} \right) \, g .
\label{eq:starproduct}
\end{align}
Let us emphasise that this star product does not correspond to the Poisson structure \eqref{eq:ClassicalPoisson}, due to the missing skew-symmetry, but is cohomologically equivalent to the Weyl-ordered one \cite{BAYEN197861, Tyutin:2001iz}. Nevertheless, as it corresponds to a constant bivector in these coordinates, it will be an associative star product.\footnote{While standard deformation quantisation associated with the Poisson structure \eqref{eq:ClassicalPoisson} (the so-called symmetric- or Weyl-ordering) yields an associative star product, it fails to induce the Clifford algebra structure required to recover the Dirac generating operator and the full kinematics of double field theory.}

Expanding the bosonic part of the exponential, one readily checks that the star-commutator reduces to
\begin{equation}
    \{ f, g \}_\star = \frac{\partial f}{\partial x^i} \star_\xi \frac{\partial g}{\partial p_i} - (-1)^{|f| |g|} \frac{\partial g}{\partial x^i} \star_\xi \frac{\partial f}{\partial p_i} + \{ f, g \}_{\star_\xi} + \mathcal{O} \Big(\hbar \partial^2_p f, \hbar \partial^2_p g ,\hbar \partial_p f \partial_p g \Big),
\label{eq:deformedPB}
\end{equation} 
where $\{ \ , \ \}_{\star_\xi}$ is the star-commutator on the degree-1 fibre (restricted to the fermionic part $\star_\xi$ of the star product), given by the standard expression \cite{BAYEN197861}
\begin{align}
    \{ f, g \}_{\star_\xi} = \hbar^{-1} \, f \, \sinh \Big( \hbar \overset{\leftarrow}{\partial} \overset{\rightarrow}{\partial} \Big) \, g = f \, \overset{\leftarrow}{\partial} \overset{\rightarrow}{\partial} \, g + \tfrac{1}{3!} \hbar^2 \, f \, (\overset{\leftarrow}{\partial} \overset{\rightarrow}{\partial})^3 \, g + \dots   \label{eq:FerminicPartDeformed} \, ,
\end{align}
with $\overset{\leftarrow}{\partial} \overset{\rightarrow}{\partial} = \overset{\leftarrow}{\frac{\partial}{\partial \xi^I}} \eta^{IJ} \overset{\rightarrow}{\frac{\partial}{\partial \xi_I}} $.
We will not need the higher order terms (quadratic in derivatives of $p$ or higher) in the following, because we will only focus on functions at most linear in the coordinate $p_i$, but they can be straightforwardly derived from the star product \eqref{eq:starproduct}.

\noindent As expected, taking the limit $\hbar \rightarrow 0$ recovers the classical Poisson bracket \eqref{eq:ClassicalPoisson},
\begin{equation}
    \lim_{\hbar \rightarrow 0} \{ f, g \}_\star = \{f, g\}.
\end{equation}
We will denote the deformation quantisation of $\mathcal{M}$ by $\mathcal{M} \longrightarrow \mathcal{M}_\hbar$. This notation is supposed to encode both the extension by the degree-2 coordinate $\hbar$ and the deformation of the graded Poisson structure. Let us mention that the extension of the QP-manifold by the degree-2 coordinate $\hbar$ can be understood as a prequantum Q-bundle \cite{Kotov:2007nr} with connection, given by $\mathcal{M}_\hbar = \mathcal{M} \times \mathbb{R}[2]$ \cite{cueca2026lecturenotessymplecticgeometry}, or in terms of graded contact Q-manifolds \cite{Mehta_2013, grabowski2013graded, contreras2026graded}.

\subsection{Derived brackets} \label{sec:derivedbrackets}

A natural question at this point is whether the Courant algebroid structures --- pairing, anchor map and Dorfman bracket --- are modified by the deformation quantisation. After all, we have replaced the Poisson bracket $\{ \cdot , \cdot \}$ with the star-commutator $\{ \cdot , \cdot \}_\star$, and the pointwise product with $\star$. One might expect $\hbar$-corrections to appear in all derived structures, e.g. in \eqref{eq:DerivedBrackets}. Remarkably, this is not the case. The Courant algebroid structures are $\hbar$-exact: they receive no corrections whatsoever, and the derived brackets survive the deformation quantisation. In this subsection, we prove this by a simple degree-counting argument.
\paragraph{Deformed derived brackets.} The natural candidates for the derived brackets after deformation quantisation are simply the star-deformed analogues of \eqref{eq:DerivedBrackets}, namely
\begin{align}
    [ e, e' ] &= - \{ \{ e, S \}_\star, e' \}_\star \, , \label{eq:DeformedDerivedBrackets1}\\
    \rho(e) f &= - \{ \{ e, S \}_\star, f \}_\star \, , \label{eq:DeformedDerivedBrackets2}\\
    \langle e, e' \rangle &= \{ e, e' \}_\star \, , \label{eq:DeformedDerivedBrackets3}
\end{align}
where $e, e' \in \Gamma (E)$ are degree-1 functions, $f \in C^\infty (M)$ is a degree-0 function, and $S$ is the degree-3 Hamiltonian. We note that, due to \eqref{eq:deg1stardefPB}, the Courant algebroid pairing \eqref{eq:DeformedDerivedBrackets3} reduces to the classical Poisson bracket, so already agrees with its classical counterpart in \eqref{eq:DerivedBrackets}. The star-commutator $\{ e, S \}_\star$ has degree $|e| + |S| -2 = 2$, so the only possible $\hbar$-contribution is of the form $\hbar \cdot C(x)$, where $C$ is a degree-0 function. Because $\hbar$ is central by construction, and the function $C$ star-commutes with degree-0 and degree-1 functions, this term does not contribute in both \eqref{eq:DeformedDerivedBrackets1} and \eqref{eq:DeformedDerivedBrackets2}. This shows that the Courant algebroid structure is rigid and survives the deformation quantisation.

\subsection{The complete tensor hierarchy}
To understand how the deformation quantisation captures the physical fields of NS-NS supergravity (in their double field theory formulation), we must examine the canonical transformations generated by functions of degree 2 in the deformed algebra\footnote{This is the first degree where a deformation appears, since $|\hbar| = 2$.}. These functions are of the form
\begin{equation}
    \alpha = v^i (x) p_i + \frac12 T_{IJ} (x) \xi^I \star \xi^J \, .
\label{eq:DegreeTwoFunction}
\end{equation}
The second term now parametrises local $\mathrm{O}(D,D) \times \mathbb{R}^+$ transformations, where $\mathrm{O}(D,D)$ comes from the skew-symmetric part $M_{IJ} = T_{[IJ]}$, while its trace generates $\mathbb{R}^+$. This agrees perfectly with the expected $\mathcal{R}_0$ of DFT, whose field content in the flux formulation consists of the generalised frame $E_A\mathstrut^M$ and the generalised dilaton $d$, respectively identified with $E_I\mathstrut^J = \exp (M_I\mathstrut^J)$ and $d = \frac12 T$ with the trace $T = T_I{}^I$. The presence of the dilaton is thus an inescapable consequence of the star product on the fibres. Because here our discussion is purely kinematical, the frames are elements of $\mathrm{O}(D,D)$ and are not required to be orthonormal (i.e. the generalised structure group is not yet reduced).
Looking at the exponential adjoint action of $\alpha$ on coordinates, we notice that only the transformation of degree-2 coordinates $p$ receives an $\hbar$-correction, when compared to \eqref{eq:pTransformationUndeformed}:
\begin{equation}
    p'_i = \frac{\partial x^j}{\partial x'^i} p_j + \frac12 \partial_i E_J\mathstrut^L E_{KL} \xi^J \xi^K + \frac12 \partial_i T \hbar \, .
\end{equation}
We now turn our attention to functions of degree $3$, which contain Hamiltonian functions. They are now of the generic form
\begin{equation}
    S = E_A\mathstrut^i (x) p_i \xi^A + \frac{1}{3!} C_{ABC}(x) \xi^A \star \xi^B \star \xi^C.
\label{eq:deg3Hamiltonian}
\end{equation}
Such Hamiltonians can also be obtained by twisting the trivial Hamiltonian $S_0 = p_i \xi^i$. When expanding the star products, the second term naturally splits in two,
\begin{equation}
\frac{1}{3!} C_{ABC} \xi^A \star \xi^B \star \xi^C = \frac{1}{3!} F_{ABC} \xi^A \xi^B \xi^C + F_A \hbar \xi^A ,
\end{equation}
where $F_{ABC} = C_{[ABC]}$ and $F_A = \frac{1}{3!} ( C_{BCA} - C_{ABC} + C_{CAB} ) \eta^{BC}$.
Imposing the deformed master equation yields:
\begin{align}
    0 &= \{S , S \}_\star \nonumber \\
    &= \eta^{AB} p_A p_B + \left( F_{BC}\mathstrut^A - 2 E_M\mathstrut^A \partial_{B} E_C\mathstrut^M \right) p_A \xi^B \xi^C + 2 \left( F^A + (\partial^B E_B\mathstrut^M) E_M\mathstrut^A \right) \hbar p_A \nonumber \\
    & \quad - \frac{1}{3} \xi^A \xi^B \xi^C \xi^D \left( \partial_{A} F_{BCD} - \frac{3}{4} {F^E}_{AB} F_{CDE} \right) \nonumber \\
    & \quad  + \hbar \xi^A \xi^B \left( \partial_C {F^C}_{AB} - 2 \partial_A F_{B} + F^C {F}_{ABC} \right) \label{eq:twoformMasterEquation}\\
    & \quad  + 2 \hbar^2 \left( \partial_A F^A +  \frac{1}{2} F_A F^A - \frac{1}{12}  F_{ABC} F^{ABC} \right),
    \label{eq:scalarMasterEquation}
\end{align}
where we defined $p_A = E_A\mathstrut^m p_m$ and $\partial_A = E_A\mathstrut^M \partial_M$. The first term implies the section condition of DFT when acting on functions $f,g$ via the derived bracket $\{ \{ \{ S, S \}_\star, f \}_\star, g  \}_\star = 0 $ \cite{Deser:2014mxa, Deser_2018}. The second and third terms give rise to the definitions of the (anchored) generalised fluxes in terms of the vielbein, and the remaining terms are the Bianchi identities that the latter must satisfy \cite{Geissbuhler:2013uka}.\footnote{In comparison to \cite{Geissbuhler:2013uka}, here $F_A$ is defined with the opposite sign.} The generalised fluxes are fixed to be
\begin{equation}
    F_{ABC} = 3 E_{[A|}\mathstrut^M \partial_M E_{|B}\mathstrut^N E_{C]N}, \qquad F_A = \partial_A T + \partial_M E_A\mathstrut^M \equiv  - 2 \partial_A d - ( \partial^B E_B\mathstrut^M ) {E_{MA}},
\label{eq:FluxesMasterEquation}
\end{equation}
where the presence of $d \in C^\infty (M)$ in $F_I$ corresponds to the freedom of adding a gradient, which vanishes under the section condition.

Let us choose the coordinate frame $E_I\mathstrut^J = \delta_I^J$, along with the following Hamiltonian function,
\begin{equation}
    S = p_I \xi^I + \frac{1}{3!} F_{IJK} \xi^I \xi^J \xi^K + \hbar F_I \xi^I \, .
\end{equation}
Then the classical master equation dictates that $F_{IJK}$ has to have $F_{ijk} \equiv H_{ijk}$ as the only non-zero component, corresponding to $H$-flux, while $F_I = (F_i, 0)$ only has a form component. These are furthermore required to be closed, via the equations $\partial_{[I} F_{JKL]} = 0$ and $\partial_{[I} F_{J]} = 0$. The Poincaré lemma guarantees that these are locally exact on $M$, so can be written as $H_{ijk} =  3 \partial_{[i} B_{jk]}$ and $F_i = 2 \partial_i d$, respectively defining the $B$-field and generalised dilaton. Globally, we observe that shifting $F_3 \equiv F_{ijk}$ and $F_1 \equiv F_i$ by these exact forms does not change the homological structure, so gauge-inequivalent deformations of the Hamiltonian are strictly classified by the first and third de Rham cohomology groups:
\begin{equation}
    \text{deformations of } \mathcal{M}_{\hbar} \quad \cong \quad H^1_\text{dR} (M) \oplus H^3_\text{dR} (M). \nonumber
\end{equation}
This provides an extension of Ševera's classification of exact Courant algebroids \cite{Severa:2017oew, Garcia-Fernandez:2016ofz, Severa:2018pag}.

Let us comment briefly about the space of degree-4 functions which are not $Q$-exact. Physically, this corresponds to violating Bianchi identities, indicating the presence of magnetic sources (e.g. NS5-branes, or exotic 7- and 9-branes \cite{Geissbuhler:2013uka, Bergshoeff:2016ncb, Bergshoeff:2019sfy}).
These are of the general form
\begin{align}
    \Omega &= W^{IJ} (x) p_I p_J + X^I\mathstrut_{JK} (x) p_I \xi^J \xi^K + Y^I (x) p_I \hbar \\
    &+ \frac{1}{4!} Z_{IJKL} (x) \xi^I \xi^J \xi^K \xi^L + \frac12 Z_{IJ} (x) \hbar \xi^I \xi^J + Z (x) \hbar^2 ,
\end{align}
where we expanded the star products. Let us mention that setting $p = 0$ is not consistent with the full $\mathrm{O}(D,D)$-covariant Q-structure, as discussed in Section~\ref{sec:reduced}.

Nevertheless, we restrict our attention to the momentum-independent subalgebra $\Cb^\infty(\mathcal{M}_\hbar) = \{ f \in C^\infty(\mathcal{M}_\hbar) \mid \partial_{p_i} f = 0 \}$ to isolate the new, physically relevant quantities, as was also done in Section~\ref{sec:reduced}. The discussion there, carries directly over to the deformed setup presented in this section.  

Formally, the emergence of the dilaton and its descendants can be understood through the Clifford algebra structure induced by the normal-ordered star product. Since $\xi^I \star \xi^J = \xi^I \xi^J + \hbar \eta^{IJ}$, the homogeneous degree-$n$ part of the projected algebra contains lower exterior-degree pieces multiplied by powers of $\hbar$:
\begin{equation}
    \Cb^\infty_n(\mathcal{M}_\hbar) \cong \bigoplus_{k=0}^{\lfloor n/2\rfloor} \hbar^k\, \Gamma\!\left(\wedge^{\,n-2k} E^*\right).
\end{equation}
Concretely, the first few sectors decompose as
\begin{align*}
    \Cb^\infty_0 \ni \lambda, \quad \Cb^\infty_1 \ni V_I \xi^I, \quad \Cb^\infty_2 \ni \tfrac{1}{2}M_{IJ} \xi^I\xi^J +  d \hbar, \quad \dots
\end{align*}
Whereas the undeformed degree-two sector yields only $\mathcal{R}_0 \cong \mathfrak{o}(D,D)$, the deformed sector yields $\mathcal{R}_0 \cong \mathfrak{o}(D,D) \oplus \mathbb{R}$, with the singlet $d \hbar$ providing the generalised dilaton. Similarly, the higher-degree sectors contain the lower-degree Clifford traces required for the DFT fluxes and Bianchi identities. Setting $\hbar=0$ reduces this structure to the undeformed tensor hierarchy of Table~\ref{tab:THwoDilaton}, while the fully deformed structure is summarised in Table~\ref{tab:THwithDilaton}.

\begin{table}[ht]
\centering
\small
\begin{align*}
{\renewcommand{\arraystretch}{1.4}
\begin{array}{c|ccc|c|c}
\text{Functions} & \multicolumn{3}{|c|}{\text{Representation}} & \text{Local expression} & \text{Physical meaning} \\ \hline \hline 
    \Cb^\infty_0(\mathcal{M}_\hbar) & \mathcal{R}_2 &=& \bullet &  \lambda(x) & \text{Gauge-for-gauge transf.} \\ \hline
    \Cb^\infty_1(\mathcal{M}_\hbar) & \mathcal{R}_1 &=& \yng(1) &  V_I(x) \xi^I & \text{Gauge transformations} \\ \hline
    \Cb^\infty_2(\mathcal{M}_\hbar) & \mathcal{R}_0 &=& \raisebox{-15pt}{\rule{0pt}{35pt}}\yng(1,1) \oplus \bullet &  \tfrac12 M_{IJ}(x) \xi^I \xi^J + d \hbar & \text{Gauge fields} \\ \hline
    \Cb^\infty_3(\mathcal{M}_\hbar) & \mathcal{R}_{-1} &=& \raisebox{-20pt}{\rule{0pt}{45pt}}\yng(1,1,1) \oplus \yng(1) &  \tfrac{1}{3!} F_{IJK}(x) \xi^I \xi^J \xi^K + F_{I}(x) \xi^I \hbar & \text{Field strengths} \\ \hline
    \Cb^\infty_4(\mathcal{M}_\hbar) & \mathcal{R}_{-2} &=& \raisebox{-25pt}{\rule{0pt}{55pt}}\yng(1,1,1,1) \oplus \yng(1,1) \oplus \bullet & \tfrac{1}{4!} Z_{IJKL} \xi^I \xi^J \xi^K \xi^L + \tfrac{1}{2} Z_{IJ} \xi^I \xi^J \hbar + Z \hbar^2 & \text{Bianchi identities \& sources} \\ \hline
    \vdots & \multicolumn{3}{c|}{\vdots} & \vdots & \vdots
\end{array}}
\end{align*}
    \caption{A tensor hierarchy associated to the deformation quantised Courant algebroid/QP-manifold $T^*[2]T[1]M$. Now, the representations correspond to the complete field content of DFT/NS-NS supergravity, including the dilaton.}
    \label{tab:THwithDilaton}
\end{table}

\subsection{Clifford algebra with Dirac generating operator}
\label{sec:CliffordDGO}
As we saw in Table~\ref{tab:THwithDilaton}, the tensor hierarchy is realised as functions $\Cb^\infty (\mathcal{M}_\hbar)$ of the coordinates $(x^i$, $\xi^I, \hbar)$, leaving out the dependence on $p_i$. In this case, one notes that the Hamiltonian does not restrict naturally to this class of functions: its adjoint action $\{ S , f \}_\star$ will generically not be in $\Cb^\infty$ . Instead, let us define a restricted differential $\Qb f = \{ S , f \}_\star\vert_{p =0}$, as in Section~\ref{sec:reduced}. For the flat Hamiltonian $S = \xi^I p_I$, this evaluates to
\begin{equation}
    \Qb = \xi^I \frac{\partial}{\partial x^I} + \hbar \eta^{IJ} \frac{\partial}{\partial \xi^I} \frac{\partial}{\partial x^J} \, .\label{eq:DefQDeformedQP}
\end{equation}
Another way to write this is
\begin{equation}
    \Qb f = \xi^I \star \frac{\partial f}{\partial x^I} \, .
\end{equation}
This form is a crucial consequence of the non-standard form of the star product \eqref{eq:starproduct} that we use. Otherwise, one would remarkably end up with the standard form of the differential:
\begin{align*}
    \Qb f &= \{S , f \}_\star \vert_{p=0} = \left(\xi^I \star \{ p_I , f \}_\star + \{ \xi^I , f \}_\star \star p_I \right)_{p=0} \\
    & = \xi^I \star \partial_I f - \{ \xi^I , \partial_I f\}_\star = \frac{1}{2} \left( \xi^I \star \partial_I f + (-1)^{|f|} \partial_I f \star \xi^I \right) \\
    &= \xi^I \partial_I f \, ,
\end{align*}
where we used the standard Moyal--Weyl star product $\star = \exp  \hbar \left( \overset{\longleftarrow}{\frac{\partial}{\partial x^I}} \overset{\longrightarrow}{\frac{\partial}{\partial p_I}} - \overset{\longleftarrow}{\frac{\partial}{\partial p_I}} \overset{\longrightarrow}{\frac{\partial}{\partial x^I}} + \overset{\longleftarrow}{\frac{\partial}{\partial \xi^I}} \eta^{IJ} \overset{\longrightarrow}{\frac{\partial}{\partial \xi^J}} \right)$.
The space of functions on the projected submanifold with coordinates $(x^i,\xi^I,\hbar)$, equipped with the differential operator $\Qb$, forms a cochain complex $\big( \Cb^\infty (\mathcal{M}_\hbar) , \Qb \big)$. Note that $\Qb$ is \emph{not} a vector field: as shown in \eqref{eq:DefQDeformedQP}, it is a second-order differential operator due to the $\hbar$-correction from the star product. Its nilpotency follows from the section condition:
\begin{equation}
   \Qb^2 = \xi^M \star \xi^N  \frac{\partial }{\partial x^M}  \frac{\partial }{\partial x^N} = (\xi^M \xi^N + \hbar \eta^{MN}) \frac{\partial }{\partial x^M}  \frac{\partial }{\partial x^N} \equiv 0 \, .
\end{equation}
This cochain complex is not obtained by a symplectic reduction of the QP2-manifold. A symplectic reduction would require a coisotropic constraint. The naive constraint $p_i \approx 0$ is not coisotropic: it would imply $\partial_i f \approx 0$ for all $f$, which would trivialise the algebra as in section~\ref{sec:reduced}. Nor can $\Qb$ be written as a Hamiltonian vector field $\{ S , \cdot \}_\star$ on the projected algebra, since the Hamiltonian $S$ depends on $p$ and therefore is not an element of $\Cb^\infty (\mathcal{M}_\hbar )$. Furthermore, due to the projection, it is not a derivation of the star product; the failure is given by
\begin{equation}
    \Phi_2 (f,g) := \Qb (f \star g ) - \Qb (f) \star g - (-1)^{|f|} f \star \Qb (g) = 2 \hbar \, \eta^{IJ} \frac{\partial f}{\partial \xi^I} \star \frac{\partial g}{\partial x^J} \, . \label{eq:Phi2}
\end{equation}
This cochain complex is not an unknown structure by any means. It is simply the Clifford algebra of $\mathrm{O}(D,D)$ generated by gamma matrices $\gamma^M$. The dictionary is the following:
\begin{align*}
    \gamma^M \quad &\longleftrightarrow \quad \xi^M \star \\
    \text{matrix multiplication} \quad &\longleftrightarrow \quad \text{star product } \star \\
    \text{Dirac generating operator } D = \gamma^M \partial_M\quad  &\longleftrightarrow \quad \Qb = \xi^M \star \partial_M \\
    \Gamma(\text{Cl}(E) ) \quad &\longleftrightarrow \quad \Cb^\infty (\mathcal{M}_\hbar),
\end{align*}
where $\text{Cl}(E)$ is the Clifford bundle of $E$.
Generically, the Dirac generating operator (DGO) does not square to zero, unless one imposes the section condition. The connection of the Clifford algebra to the generalised fluxes goes back to \cite{Geissbuhler:2013uka} and has been explored further in \cite{Hohm:2011zr, Hohm:2011dv, Jeon:2011vx, Jeon:2011sq, Jeon:2012kd, Hassler:2017yza, Carow-Watamura:2020xij}. 

The Leibniz anomaly \eqref{eq:Phi2} also measures the obstruction to a DGLA structure on the projected algebra. The star-commutator $\{ \cdot, \cdot \}_\star$ defines a graded Lie bracket on $\Cb^\infty(\mathcal{M}_\hbar)$ by associativity of $\star$. However, $\Qb$ is not a derivation of this bracket. The failure of the Leibniz rule is
\begin{equation}
    \Qb \{f,g\}_\star - \{\Qb f,g\}_\star- (-1)^{|f|}\,\{f,\Qb g\}_\star = \eta^{IJ} \left( 
    \frac{\partial f}{\partial \xi^I} \star \frac{\partial g}{\partial x^J} - (-1)^{|f||g|}\frac{\partial g}{\partial \xi^I} \star \frac{\partial f}{\partial x^J} 
    \right),
\end{equation}
which is generally non-vanishing.
This is the deformed analogue of the DGLA incompatibility discussed in Section~\ref{sec:DGLAextension}, as one sees by taking the limit $\hbar \rightarrow 0$.
Nevertheless, $\Phi_2$ is not merely an obstruction. It is precisely the correction that allows one to reconstruct the Dorfman bracket on the projected algebra. Since $S\notin \Cb^\infty(\mathcal{M}_\hbar)$, the Dorfman bracket cannot be written directly as the derived bracket $\{\{e_1,S\}_\star,e_2\}_\star$ inside the projected algebra. Instead, using $\Qb e_1=\{S,e_1\}_\star\vert_{p=0}$, one defines
\begin{equation}
[e_1,e_2]_D = \{\Qb e_1,e_2\}_\star + \frac{1}{2\hbar}\Phi_2(e_1,e_2),
\label{eq:DorfmanStarPhi2}
\end{equation}
which is well-defined because both $\Qb e_1$ and $e_2$ lie in $\Cb^\infty(\mathcal{M}_\hbar)$. In the Clifford algebra language, this reproduces the standard expression for the Dorfman bracket in terms of the Dirac generating operator \cite{Alekseev2001, Carow-Watamura:2020xij}. In this sense, the DGO expression for the Dorfman bracket is understood as the deformation quantised counterpart of the classical derived bracket construction of Roytenberg, as suggested in \cite{Bering:2006eb}.

\section{Dynamics} \label{sec:dynamics}
Having obtained the kinematics of double field theory, let us now focus on the dynamics. To do so, we need additional data in the form of a generalised metric, reducing the generalised structure group $\mathrm{O}(D,D) \times \mathbb{R}^+$ to its maximal compact subgroup $\mathrm{O}(D)_L \times \mathrm{O}(D)_R$\footnote{We will call it the double Lorentz group in the following, due to the freedom in choosing the signature of the metric on $C_\pm$.}. From the Courant algebroid perspective, this corresponds to a splitting of the generalised tangent bundle $E$ into positive and negative $D$-dimensional sub-bundles
\begin{equation}
    E = C_+ \oplus C_-,
\end{equation}
which are orthogonal with respect to the inner product on the fibre $\langle \cdot, \cdot \rangle$, and on which the latter restricts to a positive and negative definite metric respectively. Concretely, this is implemented by a constant tensor $\mathcal{H}_A\mathstrut^B \in \text{End}(E)$ satisfying
\begin{equation}
    \mathcal{H}_{AB} = \mathcal{H}_{BA}, \quad \mathcal{H}_A\mathstrut^C \mathcal{H}_C\mathstrut^B = \delta_A^B.
\end{equation}
This allows us to define projectors $P_\pm$ on the sub-bundles:
\begin{equation}
    P_\pm = \frac12 ( 1 \pm \mathcal{H} ).
\end{equation}
This operator is realised on the QP-manifold $\mathcal{M}$ via a "chiral" Euler vector field \cite{grabowski2006courant},
\begin{equation}
    \mathcal{\hat{H}} = \mathcal{H}_A\mathstrut^B \xi^A \frac{\partial}{\partial \xi^B} = \xi_+^A \frac{\partial}{\partial \xi_+^A} - \xi_-^A \frac{\partial}{\partial \xi_-^A} = \left( p_I \frac{\partial}{\partial p_I} + \xi_+^A \frac{\partial}{\partial \xi_+^A} \right) - \left( p_I \frac{\partial}{\partial p_I} + \xi_-^A \frac{\partial}{\partial \xi_-^A} \right) \equiv \mathcal{E}_+ - \mathcal{E}_-,
\end{equation}
which measures the net chirality, i.e. the difference between the number of positive and negative degree-1 coordinates $\xi_\pm^A = P_\pm (\xi^A)$. This is in contrast with the (classical) Euler vector field on $\mathcal{M}$,
\begin{equation}
    \mathcal{E} = 2 p_I \frac{\partial}{\partial p_I} + \xi^A \frac{\partial}{\partial \xi^A} = \left( p_I \frac{\partial}{\partial p_I} + \xi_+^A \frac{\partial}{\partial \xi_+^A} \right) + \left( p_I \frac{\partial}{\partial p_I} + \xi_-^A \frac{\partial}{\partial \xi_-^A} \right) \equiv \mathcal{E}_+ + \mathcal{E}_-.
\end{equation}Every function on $\mathcal{M}$ can be decomposed into eigenspaces of $\mathcal{\hat{H}}$,
\begin{equation}
    C^\infty (\mathcal{M}) = \sum_{m \in \mathbb{Z}} C^\infty_{(m)}, \qquad \mathcal{\hat{H}} (f) = m f \ \text{for} \ f \in C^\infty_{(m)}.
\end{equation}
Because the star product and hence the Poisson bracket respect this grading, the tensor hierarchy of the QP-manifold splits into definite chirality pieces.
\paragraph{Degree one:}
A function of degree one $V = V_A (x) \xi^A$ splits as
\begin{equation}
    V = V_a (x) \xi^a + V_{\Bar{a}} (x) \xi^{\Bar{a}} \equiv V_+ + V_-, \qquad \text{where} \ V_\pm \in C^\infty_{(\pm 1)} \cong \Gamma ( C_\pm) ,
\end{equation}
corresponding to the splitting of the generalised tangent bundle $E = C_+ \oplus C_-$.
\paragraph{Degree two:}
Let us now move on to degree 2, whose generic functions are of the form \eqref{eq:DegreeTwoFunction}. The part relevant to the fibre coordinates is of the form
\begin{equation}
    \alpha = \frac12 T_{AB} (x) \xi^A \star \xi^B,
\end{equation}
which expands via the star product to
\begin{equation}
    \alpha = \frac12 T_{[AB]} \xi^A \xi^B + \frac{\hbar}{2} T_A\mathstrut^A \equiv \frac12 \Lambda_{AB} \xi^A \xi^B + \frac{\hbar}{2} T_A\mathstrut^A,
\end{equation}
where $\Lambda$ is the generator of the Lie algebra $\mathfrak{o}(D,D)$ and the trace of $T$ generates an additional $\mathbb{R}$.
These generate infinitesimal canonical transformations via the adjoint action $\delta_\alpha$, which is geometrically realised as the Hamiltonian vector field
\begin{equation}
    X_\alpha = \Lambda_A\mathstrut^B \xi^A \frac{\partial}{\partial \xi^B},
\end{equation}
only depending on the $\mathfrak{o}(D,D)$ generator. 
We demand that canonical transformations preserve the eigenspaces of $\mathcal{\hat{H}}$. This amounts to requiring the vanishing of the Lie bracket of their respective vector fields:
\begin{equation}
    [X_\alpha, \mathcal{\hat{H}}] = 0,
\end{equation}
which forces the gauge parameter $\alpha \equiv \alpha_{\text{DL}}$ to decompose into purely left and right sectors:
\begin{equation}
    \alpha_{\text{DL}} = \Lambda_L + \Lambda_R = \frac12 \Lambda_{ab} (x) \xi^a \xi^b + \frac12 \Lambda_{\Bar{a} \Bar{b}} (x) \xi^{\Bar{a}} \xi^{\Bar{b}},
\end{equation}
where now $\Lambda_{ab}$ and $\Lambda_{\Bar{a} \Bar{b}}$ are generators of the Lie algebra $\mathfrak{o}(D)_L \oplus \mathfrak{o}(D)_R$. This defines the local double Lorentz group $\mathrm{O}(D)_L \times \mathrm{O}(D)_R$ as the stabiliser of the rigid splitting of the generalised tangent bundle.
\subsection{Dynamical frame and generalised metric}
To transition from this rigid structure to a dynamical geometry, we consider the generalised frame $E_A\mathstrut^I (x)$. As noted in \eqref{eq:pTransformationUndeformed}, it is generated by exponentiating the adjoint action of a generic canonical transformation $\alpha$ on the flat fibre coordinates $\xi^A$,
\begin{equation}
    \xi^I = e^{\delta_{\alpha} } \xi^A = E_A\mathstrut^I (x) \xi^A .
\end{equation}
Equipped with this frame, we construct the dynamical generalised metric by pushing forward the rigid chiral Euler vector field $\mathcal{\hat{H}}$ via the canonical transformation generated by the frame. The dynamical chiral grading operator is thus defined as:
\begin{equation}
    \mathcal{\Tilde{H}} (x) = e^{X_\alpha} \circ \mathcal{\hat{H}} \circ e^{- X_\alpha} = \mathcal{H}_I\mathstrut^J (x) \xi^I \frac{\partial}{\partial \xi^J},
\end{equation}
where $\mathcal{H}_{IJ} (x) = E_I\mathstrut^A E_J\mathstrut^B \mathcal{H}_{AB}$ is the physical generalised metric.
By construction, the local transformations $\alpha_{\text{DL}}$ are exactly the symmetries of this dynamical operator, 
\begin{equation}
    [X_{\alpha_{\text{DL}}}, \mathcal{\Tilde{H}} (x)] = 0.
\end{equation}
We have thus constructed a physical generalised metric $\mathcal{H}_{IJ}$ which is manifestly invariant under local double Lorentz transformations of the frame.
\subsection{Splitting of the Hamiltonian function}
The Hamiltonian function $S$ splits into various eigenfunctions of $\mathcal{\hat{H}}$,
\begin{equation}
    S = S_3 + S_1 + S_{-1} + S_{-3}, 
\end{equation}
where $\mathcal{\hat{H}}(S_m) = m \, S_m$. 
Explicitly, we have
\begin{align}
    S_3 &= \frac{1}{3!} F_{abc} \xi^a \xi^b \xi^c, \\
    S_1 &= \xi^a E_a\mathstrut^M p_M + \frac12 F_{ab \Bar{c}} \xi^a \xi^b \xi^{\Bar{c}} + F_a \xi^a \hbar, \\
    S_{-1} &= \xi^{\Bar{a}} E_{\Bar{a}}\mathstrut^M p_M + \frac12 F_{a \Bar{b} \Bar{c}} \xi^a \xi^{\Bar{b}} \xi^{\Bar{c}} + F_{\Bar{a}} \xi^{\Bar{a}} \hbar, \\
    S_{-3} &= \frac{1}{3!} F_{\Bar{a} \Bar{b} \Bar{c}} \xi^{\Bar{a}} \xi^{\Bar{b}} \xi^{\Bar{c}}.
\end{align}
Under local double Lorentz transformations, it transforms as
\begin{align}
    \delta_\Lambda S = \{ \Lambda_L + \Lambda_R , S \}_\star &= ( \Lambda_A\mathstrut^B E_B\mathstrut^M ) p_M \xi^A +  \frac{1}{3!} \big( 3 D_C \Lambda_{AB} + 3 \Lambda_A\mathstrut^D F_{DBC} \big) \xi^A \xi^B \xi^C \\
    &+ \big( - D^B \Lambda_{AB} + \Lambda_A\mathstrut^B F_B \big) \xi^A \hbar \\
    &\equiv ( \delta_\Lambda E_A\mathstrut^M ) p_M \xi^A + \frac{1}{3!} (\delta_\Lambda F_{ABC} ) \xi^A \xi^B \xi^C + ( \delta_\Lambda F_A ) \xi^A \hbar , 
    \label{eq:VarFrameFluxes}
\end{align}
where we introduced the flat derivative $D_A = E_A\mathstrut^M \partial_M$. This gives the transformation rules of the generalised fluxes directly from the Poisson algebra on $C^\infty (\mathcal{M}_\hbar)$, without assuming a priori that they are determined in terms of the field content.

\subsection{Action principle}
Building on this, we would like to construct an action for double field theory satisfying the following properties under the section condition:
\begin{enumerate}
    \item quadratic in derivatives and fluxes,
    \item invariant under the local subgroup $\mathrm{O}(D)_L \times \mathrm{O}(D)_R$,
    \item odd under the $\mathbb{Z}_2$ symmetry sending $\mathcal{\hat{H}} \rightarrow - \mathcal{\hat{H}}$ (reproducing the $B \rightarrow -B$ symmetry of supergravity).
\end{enumerate}
Let us start by writing down the most general scalar Lagrangian compatible with the first condition,
\begin{align}
    \mathcal{L} = c_3 \mathrm{Tr} \{ S_3 , S_3 \}_\star + c_1 \mathrm{Tr} \{ S_1 , S_1 \}_\star + c_{-1} \mathrm{Tr} \{ S_{-1}, S_{-1} \}_\star + c_{-3} \mathrm{Tr} \{ S_{-3} , S_{-3} \}_\star  \label{eq:GeneralLagrangian},
\end{align}
for constants $c_m$. Here, the trace operator $\mathrm{Tr}$ acts on a function of even degree by extracting its scalar component. We define this via the Berezin integral over the fermionic fibres, combined with the Hodge dual $*_\eta$ associated to the $\mathrm{O}(D,D)$ fibre metric $\eta$. For $f \in C^\infty_{2n} (\mathcal{M}_\hbar)$, the trace is given by
\begin{equation}\label{eq:Tr}
    \mathrm{Tr}(f) = \frac{1}{\hbar^n} \int \mathrm{d}^{2D}\xi \, *_\eta f|_{p=0} = \frac{1}{\hbar^n} \int \mathrm{d}^{2D}\xi \,  f|_{p=0} \, \xi^1 \dots \xi^{2D} \,,
\end{equation}
where the Berezin integral is normalised such that $\int \mathrm{d}^{2D}\xi \, (\xi^1 \dots \xi^{2D}) = 1$ \cite{henselder2005star}, and $|_{p=0}$ denotes the projection to $\Cb^\infty (\mathcal{M}_\hbar)$. Because the Hodge dual maps the $\xi$-independent scalar part of $f|_{p=0}$ to the top-form, the Berezin integral perfectly isolates the $\mathrm{O}(D,D)$ singlet sector, leaving a function purely on the base manifold $M$. One readily checks that this trace is cyclic ($\mathrm{Tr}(f \star g) = \mathrm{Tr} (g \star f)$), and that it reproduces the well-known trace identities of gamma matrices.

We notice that the off-diagonal terms $\mathrm{Tr} \{ S_m , S_n \}_\star$ with $m \neq n$ are not allowed due to the block-diagonal nature of the pairing $\eta$. Requiring this Lagrangian to be odd under the transformation $S_m \rightarrow S_{-m}$ forces $c_{-m} = - c_m$, leaving us with a two-parameter family of Lagrangians,
\begin{equation}
    \mathcal{L} = c_3 \left( 
    \mathrm{Tr} \{ S_3 , S_3 \}_\star - \mathrm{Tr} \{ S_{-3}, S_{-3} \}_\star 
    \right) + c_1 \left( 
    \mathrm{Tr} \{ S_1 , S_1 \}_\star - \mathrm{Tr} \{ S_{-1}, S_{-1} \}_\star
    \right).
\end{equation}
We now impose the invariance of this Lagrangian under local double Lorentz transformations,
\begin{equation}
    0 = \delta_\Lambda \mathcal{L} = 2 c_3 \mathrm{Tr} \{ \delta_\Lambda S_3 , S_3 \}_\star + 2 c_1 \mathrm{Tr} \{ \delta_\Lambda S_1 , S_1 \}_\star - 2 c_1 \mathrm{Tr} \{ \delta_\Lambda S_{-1}, S_{-1} \}_\star - 2 c_3 \mathrm{Tr} \{ \delta_\Lambda S_{-3} , S_{-3} \}_\star,
    \label{eq:VarLagrangianDL}
\end{equation}
where the variations are defined by \eqref{eq:VarFrameFluxes},
\begin{align}
    \delta_\Lambda S_3 &= \frac{1}{3!} ( \delta_\Lambda F_{abc} ) \xi^a \xi^b \xi^c , \\
    \delta_\Lambda S_1 &= (\delta_\Lambda E_a\mathstrut^M ) p_M \xi^a + \frac12 (\delta_\Lambda F_{ab \Bar{c}} ) \xi^a \xi^b \xi^{\Bar{c}} + (\delta_\Lambda F_a ) \xi^a \hbar , \\
    \delta_\Lambda S_{-1} &= (\delta_\Lambda E_{\Bar{a}}\mathstrut^M ) p_M \xi^{\Bar{a}} + \frac12 (\delta_\Lambda F_{a\Bar{b} \Bar{c}} ) \xi^a \xi^{\Bar{b}} \xi^{\Bar{c}} + (\delta_\Lambda F_{\Bar{a}} ) \xi^{\Bar{a}} \hbar , \\
    \delta_\Lambda S_{-3} &= \frac{1}{3!} ( \delta_\Lambda F_{\Bar{a}\Bar{b}\Bar{c}} ) \xi^{\Bar{a}} \xi^{\Bar{b}} \xi^{\Bar{c}} .
\end{align}
Without loss of generality, we will concentrate on the left Lorentz transformations $\delta_{\Lambda_L} \equiv \delta_L$.
The first term in \eqref{eq:VarLagrangianDL} evaluates to
\begin{equation}
    \mathrm{Tr} \{ \delta_L S_3 , S_3 \}_\star = \mathrm{Tr} \{ \frac12 D_c \Lambda_{ab} \xi^a \xi^b \xi^c , \frac{1}{3!} F_{def} \xi^d \xi^e \xi^f \}_\star = \frac12 F^{abc} D_c \Lambda_{ab}.
\end{equation}
Similarly, we can compute the variation of the second term, which results in
\begin{align}
    \mathrm{Tr} \{ \delta_L S_1 , S_1 \}_\star &= \mathrm{Tr} \{ \frac12 D_{\Bar{c}} \Lambda_{ab} \xi^a \xi^b \xi^{\Bar{c}}, \frac12 F_{de \Bar{f}} \xi^d \xi^e \xi^{\Bar{f}} \}_\star + \mathrm{Tr} \{ D^b \Lambda_{ba} \, \hbar \xi^a , F_c \, \hbar \xi^c \}_\star \\
    &= \frac12 F^{ab \Bar{c}} D_{\Bar{c}} \Lambda_{ab} - F^a D^b \Lambda_{ab},
\end{align}
while the two other terms are invariant due to skew-symmetry of the fluxes:
\begin{equation}
    \mathrm{Tr} \{ \delta_L S_{-1}, S_{-1} \}_\star = \mathrm{Tr} \{ \delta_L S_{-3}, S_{-3} \}_\star = 0.
\end{equation}
To write an action principle, we need to integrate this scalar Lagrangian against a one-density on the base manifold $M$, which in local coordinates is of the form $\mathrm{d}^D x \,  \mu(x)$ with $D = \mathrm{dim} \, M$. Such an object allows us to integrate functions on the QP-manifold $\mathcal{M}$ by extracting their $C^\infty (M)$ component via the trace operator defined previously,
\begin{equation}
    \langle f \rangle = \int_M \mathrm{d}^D x \, \mu \, \mathrm{Tr} (f).
\end{equation}
The measure is fixed by requiring the vanishing of integrals of degree-2 $Q$-exact functions,
\begin{equation}\label{eq:TrQzero}
    \langle Q(f) \rangle = 0 \, .
\end{equation}
This can be understood as a compatibility condition between the measure $\mu$ and the vector field $Q$.
Applying this to a degree-1 function $f = V_A (x) \xi^A$, where $Q (f) = - D_{A} V_{B} \, \xi^A \star \xi^B + (p_A + \hbar F_A) V^A + \tfrac12 F_{ABC} V^A \xi^B\xi^C $, one finds
\begin{align}
    \langle Q (f) \rangle &= \int_M \mathrm{d}^D x \, \mu \, \left( -D_A V^A + F_A V^A \right) \\
    &= \int_M \mathrm{d}^D x \, \mu \, \left( - \partial_M V^M + ( F_A + \partial_M E_A\mathstrut^M ) V^A \right).
\end{align}
By the master equation \eqref{eq:FluxesMasterEquation}, we have $F_A + \partial_M E_A\mathstrut^M = 2 D_A d$ for an undetermined function $d$. The goal here will be to relate it to the density function $\mu$. Integrating by parts, we get
\begin{equation}
    0 = \langle Q (f) \rangle = \int_M \mathrm{d}^D x \left( D_A \mu + 2 \mu D_A d \right) V^A .
\end{equation}
For this identity to hold for an arbitrary degree-1 function, the term inside the bracket needs to vanish. This relates the density $\mu$ to the function $d$ in the following way,
\begin{equation}
    \mu (x) = \mu_0 \, e^{-2d},
\end{equation}
where $\mu_0$ can be absorbed in the normalisation of the integral, so we will set $\mu_0 \equiv 1$ in the following. This confirms the interpretation of the generalised dilaton $d$ (or rather its exponential $e^{-2d}$) as the natural measure of double field theory. In particular, the usual integration by parts rule of DFT, namely
\begin{equation}
    \int_M \mathrm{d}^D x \, e^{-2d} D_A V^A = \int_M \mathrm{d}^D x \, e^{-2d} F_A V^A \, ,
\end{equation}
is understood as a consequence of the vanishing of $Q$-exact integrals. 
We can hence define our DFT action in the usual way,
\begin{equation}
    \mathcal{S} = \int_M \mathrm{d}^D x \, e^{-2d} \mathcal{L} \, .
\end{equation}
Coming back to the double Lorentz invariance of the Lagrangian \eqref{eq:VarLagrangianDL}, one has
\begin{equation}
    \delta_L \mathcal{L} = c_3 F_{abc} D^c \Lambda^{ab} + c_1 F_{ab \bar{c}} D^{\bar{c}} \Lambda^{ab} - 2 c_1 F_a D_b \Lambda^{ab}.
\end{equation}
Equipped with a density, we can now integrate by parts to get
\begin{equation}
    \delta_L \mathcal{L} = - c_3 (D^c - F^c) F_{abc} \, \Lambda^{ab} - c_1 ( D^{\bar{c}} - F^{\bar{c}} ) F_{ab \bar{c}} \,  \Lambda^{ab} - 2 c_1 D_a F_b \Lambda^{ab} .
\end{equation}
From the $\wedge^2 C_+ \cong \mathfrak{o}(D)_L$ component of the master equation \eqref{eq:twoformMasterEquation}, one has the Bianchi identity $-2 D_a F_b = (D^C - F^C) F_{ab C}$, simplifying the above equation further down to
\begin{equation}
    \delta_L \mathcal{L} = (c_1 - c_3 ) (D^c - F^c ) F_{abc} \Lambda^{ab}.
\end{equation}
Requiring double Lorentz invariance ($\delta_L \mathcal{L} = 0$) leads us to $c_1 = c_3$, thus singling out the unique action (up to a global factor absorbed in the normalisation),
\begin{equation}
    \mathcal{S} = \int_M \mathrm{d}^D x \, e^{-2d} \Big[ 
    \mathrm{Tr} \{ S_3 , S_3 \}_\star + \mathrm{Tr} \{ S_1 , S_1 \}_\star - \mathrm{Tr} \{ S_{-1}, S_{-1} \}_\star - \mathrm{Tr} \{ S_{-3} , S_{-3} \}_\star
    \Big].
\label{eq:ActionviaDL}
\end{equation}
We note that a specific Bianchi identity module (namely the projection of $\mathcal{R}_{-2}$ onto $\mathfrak{o}(D)_L \oplus \mathfrak{o}(D)_R$) is responsible for invariance of the Lagrangian under the local subalgebra, as was emphasized in \cite{Cederwall:2021xqi, Cederwall:2023xbj}.
Upon expansion in local coordinates, the action \eqref{eq:ActionviaDL} reduces exactly to the DFT action in the generalised flux formalism. Using the master equation, one can freely add its scalar part \eqref{eq:scalarMasterEquation} without changing the dynamics, which then reproduces the action of \cite{Geissbuhler:2013uka}. The action can be rewritten in the following suggestive way,
\begin{equation}
    \mathcal{S} = \int_M \mathrm{d}^D x \, e^{-2d} \Big(
    \sum_m \text{sgn}(m)
    \mathrm{Tr} \{ S_m , S_m \}_\star
    \Big) = \frac1\hbar \int_M \mathrm{d}^D x \, e^{-2d} \, \mathrm{Tr} \left( S \star \hat{Z} (S) \right), 
\label{eq:ActionFinalForm}
\end{equation}
where $\hat{Z}$ is the operator flipping the sign of the negative chirality eigenspaces of $\hat{\mathcal{H}}$,
\begin{equation}\label{eq:hatZ}
    \hat{Z} ( S_m ) = \mathrm{sgn} (m) \, S_m, 
\end{equation}
which can be realised on the space of degree-3 functions by the cubic polynomial operator
\begin{equation}
    \hat{Z} = \frac{1}{12} \left(  13 \hat{\mathcal{H}} - \hat{\mathcal{H}}^3 \right).
\end{equation}

\section{Generalised Cartan geometry} \label{sec:GCG}
As a second application of cochain complexes from QP-manifolds, we discuss the construction of curvature tensors in generalised Cartan geometry. For a detailed introduction to Cartan geometry and its generalisation, we refer to \cite{Hassler:2023axp,Hassler:2024hgq,Hassler:2025rag,Osten:2025lsj}.

One of the main advantages of the QP-picture of Courant algebroids is that it extends the generalised geometry tensor hierarchy to negative degrees, yielding a full cochain complex (or classical BV complex) that describes the physical degrees of freedom for NS-NS supergravity (or its duality-covariant form, namely DFT). This underlying algebraic structure provides a geometric explanation for the construction of curvature tensors in generalised Cartan geometry, which remained somewhat mysterious in \cite{Hassler:2023axp, Hassler:2024hgq, Hassler:2025rag}. These previous results in the literature would actually correspond to the tensor hierarchy of the standard, undeformed, QP-manifold (or Courant complex), as reviewed in Section~\ref{sec:QP}. We will also investigate the deformation quantisation of this setting that introduces several new features --- model algebra components, torsion and curvature corresponding to the dilaton as they were anticipated in \cite{Butter:2022iza}.

\paragraph{Cartan Geometry.} Cartan geometry may be understood as a curved version of Klein geometry and provides a common language for several familiar geometric structures, including Riemannian, projective, and conformal geometries \cite{Sharpe:1997,cap2009parabolic}. Rather than taking a vector space as the local description, as in ordinary Riemannian geometry, one starts from a homogeneous space $G/H$, where $G$ is a Lie group and $H$ is a closed subgroup. A general manifold is then treated as locally resembling this homogeneous model, but only in an infinitesimal sense: curvature measures precisely the failure of these local Klein models to assemble into a globally homogeneous space. In the relativistic case, the flat model is Minkowski space, $\mathrm{ISO}(1,D-1)/\mathrm{SO}(1,D-1) \cong \mathbb{R}^{1,D-1}$, and the Cartan connection packages the vielbein, associated with translations, together with the spin connection, associated with Lorentz rotations, into one unified object. Let $M$ be a manifold of dimension $D$, and let $\mathfrak{g}$ and $\mathfrak{h}$ be the Lie algebras corresponding to $G$ and $H$. We assume that the model space has the correct dimension, namely $\dim(\mathfrak{g}/\mathfrak{h}) = D$. A \emph{Cartan geometry} modelled on $G/H$ consists of a principal $H$-bundle $\pi: P \to M$ equipped with a $\mathfrak{g}$-valued one-form $\theta$, called the \emph{Cartan connection}. This form is required to satisfy the following three conditions: 
\begin{enumerate} 
    \item \textit{Pointwise identification with the model algebra:} for each $p\in P$, the map $\theta |_p : T_p P \to \mathfrak{g}$ is a linear isomorphism;
    \item \textit{$H$-equivariance}: Compatibility with the right $H$-action: for every $h\in H$, one has $R_h^* \theta = \mathrm{Ad}_{h^{-1}}\theta$. An alternative phrasing of this is that $\theta$ and all its constituents transform covariantly under $H$-gauge transformations.
    \item There is no geometric information in the gauge direction, i.e. $\xi \in \mathfrak{h}$ is identified with the corresponding vector fields $X_\xi$ in $P$ via the Cartan connection, $\theta(X_\xi) = \xi$. \end{enumerate} 
The curvature of the Cartan connection is the $\mathfrak{g}$-valued two-form 
\begin{equation} 
    \Theta = - \mathrm{d}\theta + \tfrac12 [\theta,\theta] .
\end{equation} 
A particularly important class is given by \emph{reductive} Cartan geometries. In this case the Lie algebra admits an $\mathrm{Ad}(H)$-invariant decomposition $\mathfrak{g} = \mathfrak{h} \oplus \mathfrak{g}/\mathfrak{h}$, so that the Cartan connection can be written as $\theta = \omega + e$. The component $\omega$ is an ordinary Ehresmann connection, such as the spin connection in gravitational applications, while $e$ identifies tangent directions on $M$ with the model directions $\mathfrak{g}/\mathfrak{h}$ and therefore plays the role of a (co)frame. Under the same splitting, the Cartan curvature separates into two natural pieces: an $\mathfrak{h}$-valued curvature $R$ and a $\mathfrak{g}/\mathfrak{h}$-valued torsion $T$. 

\paragraph{Generalised Cartan geometry.} The idea of Cartan geometry can be adapted to the setting of generalised geometry, leading to what is known as \emph{generalised Cartan geometry} \cite{Hassler:2024hgq}. This formulation is closely related to earlier work of Pol\'a\v{c}ek and Siegel \cite{Polacek:2013nla}, where Cartan-geometric methods were used to give a systematic construction of covariant torsion and curvature tensors in generalised geometry and double field theory. The key modification is that the usual absolute parallelism of Cartan geometry is replaced by an isomorphism involving the doubled tangent space: 
\begin{equation} \theta |_p : T_p P \oplus T_p^* P \longrightarrow \mathfrak{d}.
\end{equation} 
Here, $\mathfrak{d}$ is a Lie algebra of dimension $2(D+\dim \mathfrak{h})$ equipped with a non-degenerate split symmetric ad-invariant bilinear form $\eta$. The subalgebra $\mathfrak{h} \subset \mathfrak{d}$ is assumed to be isotropic, and the connection is required to preserve the pairing $\eta$.

\subsection{The underlying QP-manifold}
Cartan geometry is based on the geometry of a principal $H$-bundle $P \rightarrow M$, where locally $P \sim M \times H$. Consequently, generalised Cartan geometry will be based on the generalised geometry or the Courant algebroid of $P$. In graded differential geometry terms, this means the underlying object is $\mathcal{P} = T^*[2] T[1] P$ instead of $\mathcal{M} = T^*[2] T[1] M$. As coordinates on a patch of $\mathcal{P}$, we fix
\begin{equation}
    x^{\mathcal{M}} = \begin{pmatrix} y^\mu & x^M & 0 \end{pmatrix}\,, \qquad
    \xi^{\mathcal{A}} = \begin{pmatrix} \xi^\alpha & \xi^A & \xi_\alpha \end{pmatrix}\,, \qquad \text{and} \qquad
    p_{\mathcal{M}} = \begin{pmatrix} p_\mu & p_M & 0 \end{pmatrix}\, ,
\end{equation}
where the $y^{\mu}$ are coordinates on $H$. We work on a solution of the section condition for general functions $f,g$ on $M$
\begin{equation}
    \eta^{MN} \frac{\partial f}{\partial x^M} \frac{\partial g}{\partial x^N} = 0, \qquad 0= p_M p_N \eta^{MN}.
\end{equation}
The standard solution $x^M = (x^m , 0)$ and $p_M = (p_m , 0)$ will reproduce the notation from sections~\ref{sec:QP} and~\ref{sec:Kinematics}.

We introduce the canonical degree-2 QP-manifold\footnote{Any other degree would be also possible, e.g. $T^*[n]T[1]H$. The form of the Hamiltonian would not change.} for a Lie group $H$ as $T^*[2]T[1]H$:
\begin{center}
    \begin{tabular}{l|cccc}
        grading & 0 & \multicolumn{2}{c}{\quad 1 \quad} & 2 \\ \hline
        coordinates \quad  & \quad  $y^\mu$ & \quad $\xi^\alpha$ \quad & \quad $\xi_\alpha$ \quad &  \quad $p_\mu$
    \end{tabular}
\end{center}
$\xi_\alpha$ are the degree shifted Lie algebra generators with Poisson duals $\xi^\alpha$, $\{ \xi_\alpha , \xi^\beta \} = \delta_\alpha^\beta$, together with the coordinates $x^\alpha$ and their canonical dual momenta $p_\alpha$, $\{y^\alpha , p_\beta \} =\delta_\beta^\alpha$.

Normally, this is overkill: a Q-manifold alone already incorporates the Chevalley-Eilenberg complex and the full cohomological structure of a Lie algebra, without the need for a P-structure. Nevertheless, transitioning to the QP-picture allows the Lie algebra structure to be explicitly encoded in the Hamiltonian. In coordinates adapted to $H$ via a left-invariant frame $p_\alpha = {e_\alpha}^\mu p_\mu$, this reads:
\begin{equation}
    S_H = \xi^\alpha p_\alpha + \frac12 {f^\alpha}_{\beta \gamma} \xi_\alpha \xi^\beta \xi^\gamma .
\end{equation}
The master equation $\{S_H , S_H \} =0$ is a consequence of the Jacobi identity of the Lie algebra $\mathfrak{h}$. This Hamiltonian will also reproduce the 'covariant derivative' when acting on tensors $T = {T_{\alpha \dots}}^{\beta \dots} \xi^{\alpha} \dots \xi_\beta \dots$\,.

Formally, we now also perform the star-deformation of this manifold exactly as before:
\begin{equation}
    \mathcal{P} =  T^* [2] T[1] P \quad \rightarrow \quad \mathcal{P}_\hbar
\end{equation}
The relevant coordinates are organised as follows:
\begin{center}
    \begin{tabular}{l|c|ccc|c}
        grading & \quad 0\quad & \multicolumn{3}{c|}{\quad 1 \quad} & \quad 2 \quad  \\ \hline
        coordinates \quad  & \quad$y^\mu , x^M $\quad & \quad $\xi^\alpha$ \quad & \quad $\xi^M$ \quad & \quad $\xi_\alpha$ \quad & \quad $p_\mu , p_M,\hbar$ \quad
    \end{tabular}
\end{center}We equip this space with the deformed Poisson bracket \eqref{eq:deformedPB}. The $\mathfrak{h}$-sector is similarly deformed, though this only becomes relevant for quantities containing cubic terms in both $\xi_\alpha$ and $\xi^\alpha$. Note that deformation quantisation must be performed on the complete extended QP-manifold; one cannot restrict the deformation to the base $T^*[2] T[1] M$. However, on this full space, it takes the canonical form.

\subsection{Introducing the model algebra}
The combined Hamiltonian
\begin{align}
    \tilde{S} &= S_M + S_H \\
    & = \xi^A p_A + \xi^\alpha p_\alpha + \frac{1}{2} {f^\alpha}_{\beta \gamma} \xi_\alpha \xi^\beta \xi^\gamma \equiv \xi^{\mathcal A} p_{\mathcal{A}} + \frac{1}{2} {f^\alpha}_{\beta \gamma} \xi_\alpha \xi^\beta \xi^\gamma
\end{align}
trivially satisfies the master equation. As in standard generalised geometry on $M$, one can twist the setup by introducing a frame $E_{\mathcal{A}}{}^{\mathcal{M}}$ on the fibres of the bundle generated by $\xi_{\mathcal{M}}$.  In this case $p_\mathcal{A} = {E_\mathcal{A}}^{\mathcal{M}} p_{\mathcal{M}}$ should be understood as a twist\footnote{The condition on this twist is that the resulting structure constants $f_{\alpha \mathcal{B}}{}^{\mathcal{C}}$ are constant. With that, any explicit dependence on $y$ drops out.} of $p_\mathcal{M} = ( p_\mu , p_i , p^i , p^\mu ) \equiv ( p_\mu , p_i , 0 , 0 )$. The case of primary interest here, namely the one corresponding to the introduction of a model algebra, arises from a particular extended generalised frame ${E_\mathcal{A}}^{\mathcal{M}}(y)$. Under this twist, the Hamiltonian takes the form
\begin{align}
    \tilde{S} &= \xi^{\mathcal A} p_{\mathcal{A}} + \frac{1}{2} f_{\alpha \mathcal{B}}{}^{\mathcal{C}} \xi^\alpha \star \xi^{\mathcal{B}} \star \xi_{\mathcal{C}} \, ,
\end{align}
with structure constants $f_{\alpha \mathcal{B}\mathcal{C}}$ corresponding to the model algebra. The only additional constraint is that $f_{\alpha \beta \gamma} = 0$, ensuring that $\mathfrak{h}$ is a true subalgebra of $\mathfrak{d}$.

A particularly instructive example is the following Hamiltonian\footnote{In particular, we will ignore a contribution of the form $f_M \xi^M \hbar$. This is not part of the model algebra, but rather is the physical characterisation of the background (corresponding to the dilaton flux). Furthermore, such a term could not be produced from a twist of the form ${E_\mathcal{A}}^{\mathcal{M}}(y)$.}: 
\begin{equation}
    \tilde{S} = \xi^{\mathcal A} p_{\mathcal{A}} + \frac{1}{2} {f^\alpha}_{\beta \gamma} \xi_\alpha \xi^\beta \xi^\gamma + \frac{1}{2} f_{\alpha MN} \xi^\alpha \xi^M \xi^N + f_\alpha \xi^\alpha \hbar.\label{eq:DifferentialGeneralisedPoincare}
\end{equation}
The master equation for this Hamiltonian is satisfied, provided the structure constants $f_{\alpha BC}$ define an action of $\mathfrak{h}$ on the generalised tangent bundle fibres:
\begin{equation}
    2 f_{\alpha B}{}^D f_{\beta D}{}^C = f^\gamma{}_{\alpha \beta}f_{\gamma B}{}^C, \quad f^{\alpha}{}_{\beta \gamma} f_\alpha = 0.
\end{equation}
This is what was described as a model algebra $\mathfrak{d}$ of 'generalised Poincaré type' in \cite{Hassler:2024hgq}. We will compute the generalised Cartan geometry for this specific case in the next section. The calculations for the most general case proceed identically, albeit yielding more cumbersome expressions.

\subsection{The hierarchy of generalised Cartan geometry}
The structures of generalised Cartan geometry will then emerge from a similar reduction to the Clifford algebra bundle. Just as in the standard formulation of generalised fluxes, Bianchi identities in double field theory discussed above, the construction of Cartan curvatures will have the natural structure of a cochain complex (which in some sense could be understood as its classical BV complex). In order to reconstruct and extend the results from \cite{Polacek:2013nla,Hassler:2024hgq}, we make the following definitions:
\begin{itemize}
    \item In addition to projecting onto functions $\Cb^\infty(\mathcal{P}_\hbar)$ that are independent of the momenta, generalised Cartan geometry requires restricting to a \textit{parabolic subalgebra}:
    \begin{equation}
        \Cb^\infty_{+}(\mathcal{P}_\hbar) = \{ f \in C^\infty(\mathcal{P}_\hbar) \ \vert \ \partial_{\xi^\alpha} f = 0\ \& \ \partial_{y^\mu} f = 0 \}\,.
    \end{equation}
    This also implies that physical functions do not depend on the auxiliary coordinates $y^\mu$.
    
    In Table \ref{tab:CartanGeometry} we summarise our conventions and the expansions for the physical objects.
    
    \item The differential will be defined as the projection $\tilde{Q} = \{ \tilde{S} , \ \cdot \ \}_\star \vert_{p = 0}$. Note that this differential can generate terms outside of $\Cb_+^\infty (\mathcal{P}_\hbar)$.
    
    The operator $\tilde{Q}$ combines the differential generated by $S_M \in C^\infty (\mathcal{M}_\hbar)$, that was discussed in earlier sections, with the action of $S_H$ on the model algebra.
\end{itemize} 

\begin{table}[ht]
\centering
\begin{footnotesize}
\begin{align*}
{\renewcommand{\arraystretch}{1.4}
\begin{array}{c|c|c|c|c}
\text{functions} & \multicolumn{2}{|c|}{\text{representation}}  & \multicolumn{2}{|c}{\text{physical interpretation}} \\ \hline \hline 
    \Cb^\infty_{0+}(\mathcal{P}_\hbar) & f \in \mathcal{R}_2 &  f(x) & & \text{gauge for gauge} \\ \hline
    \multirow{2}{*}{$\Cb^\infty_{1+}(\mathcal{P}_\hbar)$} & \multirow{2}{*}{$\Lambda \in \mathcal{R}_1$} &  V^I(x) \xi_I & \text{generalised diffeomorphisms} & \multirow{2}{*}{\text{gauge}} \\
    & & + \Lambda^\alpha(x) \xi_\alpha & \mathfrak{h}\text{-gauge transformations} \\ \hline
    \multirow{2}{*}{$\Cb^\infty_{2+}(\mathcal{P}_\hbar)$} & \multirow{2}{*}{$\theta \in \mathcal{R}_0$}  &  M_{IJ}(x) \xi^I \xi^J + d(x) \hbar & \text{frame and dilaton} & \multirow{2}{*}{\text{gen. Cartan connection}} \\ 
    & & + \Omega^\alpha_I(x) \xi_\alpha \xi^I + \rho^{\alpha \beta}(x) \xi_\alpha \xi_\beta & \text{hierarchy of connections} \\ \hline
    \multirow{3}{*}{$\Cb^\infty_{3+}(\mathcal{P}_\hbar)$} & \multirow{3}{*}{$ \Theta \in \mathcal{R}_{-1}$} &  T_{IJK}(x) \xi^I \xi^J \xi^K + T^\alpha(x) \xi_\alpha \hbar & \text{generalised torsions}  & \multirow{3}{*}{\text{gen. Cartan curvature}} \\ 
    & & + R_{IJ}^\alpha(x) \xi_\alpha \xi^I \xi^J + R_I^{\alpha \beta}(x) \xi_\alpha \xi_\beta \xi^I & \text{hierarchy of curvatures} & \\ 
    & & + R_I(x) \xi^I \hbar  & \text{divergence} & \\ \hline
    \multirow{3}{*}{$\Cb^\infty_{4+}(\mathcal{P}_\hbar)$} & \multirow{3}{*}{$Z \in \mathcal{R}_{-2}$} &  Z_{IJKL}(x) \xi^I \xi^J \xi^K \xi^L & \text{Bianchi identity for $T_{IJK}$} & \multirow{3}{*}{\text{Bianchi identities}} \\
    & & +Z_{IJK}{}^\alpha (x) \xi^I \xi^J \xi^K \xi_\alpha & \text{Bianchi identity for ${R_{MN}^\alpha}$} & \\
    & & \vdots & \vdots & \\ \hline
    \vdots & \vdots & \vdots & & 
\end{array}}
\end{align*}
\end{footnotesize}
    \caption{A tensor hierarchy for generalised Cartan geometry, corresponding to functions $\bar{C}^\infty_{n+}(\mathcal{P})$ on the deformation quantised $T^*[2]T[1]P$.}
    \label{tab:CartanGeometry}
\end{table}
For example, the generalised Cartan connection $\theta$ will be the most general object of degree 2, non-negative $\mathfrak{h}$-degree and independent of momenta:
\begin{equation}
    \theta =  M_{IJ}(x) \xi^I \xi^J + d(x) \hbar + \Omega_I^\alpha(x) \xi_\alpha \xi^I + \rho^{\alpha \beta}(x) \xi_\alpha \xi_\beta
\end{equation}
As in the standard metric or flux formulation of DFT, the curvature is identified as the deviation of the differential twisted by a degree-2 object -- here, the generalised Cartan connection. This defines the \textit{generalised Cartan curvature} as
\begin{align}
    \Theta = \Delta_\theta \tilde{S} &= e^{\delta_\theta} \tilde{S} - \tilde{S} = \{ \tilde{S} , \theta \}_\star + \frac12 \{ \{ \tilde{S} , \theta \}_\star , \theta \}_\star + \ldots = \tilde{Q} \theta + \mathcal{O}(\theta^2)\,. \label{eq:NonLinear}
\end{align}
Using identity \eqref{eq:canonicaltr} it is easy to see that $\Theta$ satisfies the (non-linear) Bianchi identities:
\begin{equation}
    \tilde Q \Theta + \tfrac12 \{ \Theta, \Theta \}_\star = 0.
\end{equation}
This way, it is obvious that the generalised Cartan curvature is just the Maurer--Cartan form in the framework of the DGLA, a construction already considered in \cite{Greitz:2013pua}. In an abstract way, identifying $e^{\delta_\theta}$ with a group element, this can also be understood as
\begin{equation}
     \{ \Theta, \, \cdot \, \}_\star = e^{\delta_\theta} \tilde{Q} e^{-\delta_\theta}  \,.
\end{equation}
This motivates the following cochain complex that describes \textit{linearised} generalised Cartan geometry:
\begin{equation}
    f \overset{\tilde{Q}}{\longrightarrow} \Lambda \overset{\tilde{Q}}{\longrightarrow} \theta \overset{\tilde{Q}}{\longrightarrow} \Theta \overset{\tilde{Q}}{\longrightarrow} Z \overset{\tilde{Q}}{\longrightarrow} \ldots \,. \label{eq:CartanComplex}
\end{equation}
We refer to Table \ref{tab:CartanGeometry} for the relevant expansions. Using these expansions alongside the differential derived from \eqref{eq:DifferentialGeneralisedPoincare}, one naturally reproduces the (linearised) curvatures $R$ and torsion-like objects\footnote{These are defined as objects that do not contain derivatives of the connections.} $T$ initially derived in \cite{Polacek:2013nla,Hassler:2024hgq}: 
\begin{align}
    S_{KLM} &= T_{KLM} = f_{\alpha [ KL} \Omega_{M]}^\alpha \nonumber \\
    S_{KL}^\alpha &= R_{KL}^\alpha = 2 \partial_{[K} \Omega_{L]}^\alpha + \rho^\alpha f_{\alpha KL} \nonumber \\
    S_M^{\alpha \beta} &= R_M^{\alpha\beta} = - \partial_M  \rho^{\alpha \beta} \label{eq:linearisedCurvatures} \\
    S_M &= R_M = \partial_M d + f_\alpha \Omega^\alpha_M \nonumber \\
    S^\alpha &= T^\alpha = \rho^{\alpha \beta} f_\beta \, , \nonumber
\end{align}
together with an extension including the dilaton and the trace model algebra $f_\alpha$. As in both the undeformed QP-case and the usual generalised Cartan geometry, there are two main ways to treat the $M_{IJ}$ component:
\begin{itemize}
    \item \textbf{Generalised metric formalism:} Here, the connection $\Omega_M^\alpha$ provides the fundamental description of the geometry --- with the final aim as in \cite{Hohm:2011si} to fix the connection in terms of the generalised metric and dilaton. The torsion-free condition directly imposes constraints on $\Omega_M^\alpha$, as reviewed in detail in \cite{Hassler:2024hgq}. For this the fields $M_{IJ}$ are chosen to be zero or constant.
    
    \item \textbf{Generalised flux formalism:} Alternatively, one can treat the frame $E_I{}^A$ (the integrated form of $M_{IJ}$\footnote{The diagonal part of the generalised Cartan connection can be exponentiated separately.}) as the fundamental physical field. This leads to:
    \begin{align}
    S_{KLM} &= T_{KLM} = f_{\alpha [ KL} \Omega_{M]}^\alpha - F_{KLM} \nonumber \\
    S_{KL}^\alpha &= R_{KL}^\alpha = 2 \partial_{[K} \Omega_{L]}^\alpha + \rho^\alpha f_{\alpha KL} + F^{M}{}_{KL} \Omega_M^\alpha \nonumber \\
    S_M^{\alpha \beta} &= R_M^{\alpha\beta} = - \partial_M  \rho^{\alpha \beta} \label{eq:linearisedCurvaturesFluxes} \\
    S_M &= R_M = \partial_M d + f_\alpha \Omega^\alpha_M \nonumber \\
    S^\alpha &= T^\alpha = \rho^{\alpha \beta} f_\beta \nonumber
\end{align}
with the generalised fluxes $F_{KLM} = 3 (\partial_{[K} E_L{}^A) E_{M]A}$. The torsion-free condition then fixes the connection entirely in terms of the generalised frame. Let us note that for the complex $\Cb^\infty_+ (\mathcal{P}_\hbar)$, as discussed in Section \ref{sec:reduced}, $\tilde{Q}$ is only homological for constant generalised fluxes. In the general case, one can still resort to the complete QP-manifold complex $C^\infty (\mathcal{P}_\hbar)$, as discussed in Section \ref{sec:DGLAextension} -- the only drawback being that this complex contains additional non-physical fields. In the linearised theory, the generalised fluxes would vanish.
\end{itemize}
A similar analysis can be systematically carried out for all higher stages of the generalised Cartan geometry complex \eqref{eq:CartanComplex}.

\section{Conclusion and outlook} \label{sec:Outlook}
We have demonstrated that formal deformation quantisation of degree-2 QP-manifolds naturally bridges two distinct approaches to double field theory (DFT): graded symplectic geometry and the Clifford algebra formalism of Dirac generating operators \cite{Alekseev2001, Severa:2018pag, Carow-Watamura:2020xij}. By equipping the fermionic fibres with a normal-ordered Moyal--Weyl star product, their exterior algebra is deformed into a Clifford algebra. This procedure systematically completes the classical tensor hierarchy down to the negative-degree representations ($\mathcal{R}_{n \leq 0}$). Crucially, this resolves the geometric origin of the generalised dilaton. While classical degree-2 generators take values strictly in $\mathfrak{o}(D,D)$, the non-commutative deformation forces them to acquire a trace proportional to the formal parameter $\hbar$. This trace identifies directly with the generalised dilaton field, naturally extending the duality structure group from $\mathrm{O}(D,D)$ to $\mathrm{O}(D,D) \times \mathbb{R}^+$.

This deformation, followed by the restriction to the physical fields, fundamentally alters the algebraic structure of the hierarchy. Before projection, the full algebra of functions on the deformed manifold actually remains a strict differential graded Lie algebra (DGLA). Since the full star product is associative, the deformed differential $Q = \{S, \cdot\}_\star$ acts as a perfect inner derivation of both the star product and the star-commutator. However, capturing the physical fields requires restricting to the momentum-independent subspace ($p=0$). Because the Hamiltonian $S$ inherently depends on $p$, projecting to this physical slice removes the generator of the differential from the algebra itself. Consequently, the projected differential can no longer be expressed as an adjoint action, and its compatibility with the bracket is broken. While the restricted space retains a graded Lie algebra structure via the star-commutator, the failure of the Leibniz rule means the DGLA structure is lost. Instead, the kinematics are governed by a cochain complex equipped with an associative star product, where the differential is realised as the Dirac generating operator acting on the Clifford bundle.

This framework cleanly separates kinematics from dynamics. The deformed master equation $\{S, S\}_\star = 0$ purely governs the kinematic data, yielding the section condition and the Bianchi identities for the DFT fluxes. The dynamics, however, emerge from an independent action principle. Introducing a physical generalised metric defines a chiral grading operator that splits the Hamiltonian into left and right eigenspaces. Demanding local double Lorentz invariance and a $\mathbb{Z}_2$ symmetry exchanging the chiral sectors uniquely fixes the DFT action. Furthermore, this same cochain complex supplies the precise algebraic data needed to construct covariant curvature and torsion tensors in generalised Cartan geometry.

\subsection*{Future directions}

The framework developed here points to several natural avenues for further investigation:

\paragraph{Algebraic structure and link to BV formalism.} The failure of the projected differential to act as a derivation of the star-commutator deserves closer scrutiny. On the projected subalgebra $\Cb^\infty(\mathcal{M}_\hbar)$, the differential $\Qb$ is no longer an inner derivation, and the Leibniz rule fails by a term controlled by the odd symplectic structure $\eta_{IJ}\,dx^I \wedge d\xi^J$ on the $(x,\xi, \hbar)$ submanifold. This is suggestive of a (non-commutative) Batalin--Vilkovisky (BV) structure, in which the second order part of $Q$ plays the role of a BV Laplacian \cite{Bering:2006eb}. Indeed, the expansion of the deformed master equation, $\{S,S\}_\star = \{S,S\}_{\mathrm{cl}} + \hbar\Delta S + \dots$, takes the form of a quantum master equation. Studying the algebraic properties of this projected structure, and understanding its possible relation to non-commutative BV algebras \cite{Bering:2006eb} remains an open problem.

\paragraph{Ramond--Ramond (RR) fields and extended objects.} The Clifford algebra structure is inherently suited to capture the RR-sector. In the $\mathrm{O}(D,D)$ setting, RR fields can be interpreted as $\mathrm{O}(D,D)$ spinors within a Fock space embedded in $\Cb^\infty(\mathcal{M}_\hbar)$, with the action principle~\eqref{eq:ActionFinalForm} expected to reproduce the Mukai pairing once the Hamiltonian function is replaced with the RR field strength polyform \cite{Coimbra:2011nw, Hohm:2011zr, Hohm:2011dv, Jeon:2011vx, Jeon:2011sq, Jeon:2012kd, Geissbuhler:2013uka, Severa:2018pag}. In this setting, the negative-degree representations $\mathcal{R}_{n \leq 0}$ are simple enough ($\mathcal{R}_0$ contains the frame and dilaton, $\mathcal{R}_{-1}$ the three-form and one-form fluxes), and the undeformed Hamiltonian admits deformations parametrised by $H^1_{\mathrm{dR}}(M) \oplus H^3_{\mathrm{dR}}(M)$, corresponding to the dilaton flux and the NS-NS $H$-flux respectively. These cohomology classes have a direct physical interpretation in terms of extended objects: non-trivial classes in $H^3$ correspond to the presence of NS5-branes (the magnetic sources of H-flux), while non-trivial classes in $H^1$ correspond to codimension-two exotic 'NS7-branes' (the magnetic sources of the dilaton flux). Furthermore, the space of Bianchi identities $\mathcal{R}_{-2}$ also contains a singlet part, potentially signalling the presence of exotic 'NS9-branes'. The negative-degree tensor hierarchy---which our deformation makes manifest---contains precisely the dual potentials that couple to these extended objects \cite{Geissbuhler:2013uka, Bergshoeff:2016ncb, Bergshoeff:2019sfy}. 

\paragraph{Exceptional generalised geometry.} Alternatively, one could attempt a direct generalisation to exceptional field theory (ExFT) \cite{Hohm:2013pua, Hohm:2013vpa, Hohm:2013uia, Hohm:2014fxa}, where the RR fields are part of the duality-covariant multiplet from the start. However, the exceptional case is substantially more involved: exceptional groups introduce non-trivial representation structure already at positive degrees $\mathcal{R}_2, \mathcal{R}_3, \dots$ Standard QP-manifold approaches to exceptional geometry struggle to reach these negative-degree representations and depend heavily on the degree of the underlying manifold \cite{Arvanitakis:2018cyo, Osten:2023iwc}. The Q-manifold perspective developed here offers a degree-independent alternative: one can define a graded manifold directly using the representations of the tensor hierarchy, with coordinates $x^{I_1}$, $\xi^{I_1}$, $\xi^{I_2}, \dots$ corresponding to each level of the hierarchy. Following the notation of \cite{Osten:2024mjt, Hassler:2025rag}, the homological operator would take the form
\begin{equation}
Q = \xi^{I_1} \frac{\partial}{\partial x^{I_1}} + \sum_{p \geq 1} D_{M_{p+1}}{}^{K_p L_1} \xi^{M_{p+1}} \frac{\partial}{\partial \xi^{K_p}} \frac{\partial}{\partial x^{L_1}},
\end{equation}
where the $D$-symbols in the second term encode the tensor hierarchy differential $\hat{\partial}$. The condition $Q^2 = 0$ then encodes both the ExFT section condition and the nilpotency of $\hat{\partial}$. A deformation quantisation of this exceptional Q-manifold could provide the missing negative-degree sector, but the technical complexity is significantly greater than in the $\mathrm{O}(D,D)$ case.

\paragraph{Worldsheet interpretation.} It would be desirable to understand the deformed hierarchy from the perspective of the worldsheet. One route is through graded contact geometry: adjoining $\hbar$ to the QP2-manifold can be viewed as a 'contactification', yielding a graded contact Q-manifold \cite{Mehta_2013, grabowski2013graded}. The AKSZ-contact formalism \cite{contreras2026graded} then produces sigma-models whose target-space structure is governed by Courant--Jacobi algebroids, in much the same way that the Courant sigma-model is the AKSZ theory associated to a Courant algebroid \cite{Ikeda:2002wh, Roytenberg:2006qz}. A complementary route passes through current algebras. Alekseev and Strobl associated a current algebra to any Courant algebroid \cite{Alekseev:2004np}, a construction extended to brane currents via QP-manifolds \cite{Ikeda:2011ax} and related to the tensor hierarchy in \cite{Arvanitakis:2021wkt}. Since our deformation quantises precisely the Poisson algebra underlying both constructions, a natural question is whether these worldsheet structures deform along with it, perhaps along the lines of \cite{Basile:2019pic}.

\subsection*{Acknowledgements}

We are particularly grateful to Maxim Grigoriev for pointing out to us the relation between Clifford algebras and deformation quantisation, which provided the starting point for this project. Furthermore, we wish to thank Thanasis Chatzistavrakidis, Marija Dimitrijević Ćirić and Jakob Palmkvist for useful comments and discussions. F.~H.~and A.~S.~are supported by the SONATA BIS grant 2021/42/E/ST2/00304 from the National Science Centre (NCN), Poland. The research of D.~O.~was part of the SONATA grant No.~2024/55/D/ST2/01205 funded by NCN.

\bibliographystyle{jhep}
\bibliography{literature}

@article{Bonezzi:2019bek,
    author = "Bonezzi, Roberto and Hohm, Olaf",
    title = "{Duality Hierarchies and Differential Graded Lie Algebras}",
    eprint = "1910.10399",
    archivePrefix = "arXiv",
    primaryClass = "hep-th",
    doi = "10.1007/s00220-021-03973-8",
    journal = "Commun. Math. Phys.",
    volume = "382",
    number = "1",
    pages = "277--315",
    year = "2021"
}

@article{Siegel:1993xq,
    author = "Siegel, W.",
    title = "{Two vierbein formalism for string inspired axionic gravity}",
    eprint = "hep-th/9302036",
    archivePrefix = "arXiv",
    reportNumber = "ITP-SB-93-2",
    doi = "10.1103/PhysRevD.47.5453",
    journal = "Phys. Rev. D",
    volume = "47",
    pages = "5453--5459",
    year = "1993"
}

@article{Siegel:1993th,
    author = {Siegel, W.},
    journal = {Phys. Rev. D},
    title = {{Superspace duality in low-energy superstrings}},
    year = {1993},
    pages = {2826--2837},
    volume = {48},
    number = {48},
    archiveprefix = {arXiv},
    doi = {10.1103/PhysRevD.48.2826},
    eprint = {hep-th/9305073},
    reportnumber = {ITP-SB-93-28},
}

@inproceedings{Siegel:1993bj,
    author = "Siegel, W.",
    title = "{Manifest duality in low-energy superstrings}",
    booktitle = "{International Conference on Strings 93}",
    eprint = "hep-th/9308133",
    archivePrefix = "arXiv",
    reportNumber = "ITP-SB-93-50",
    month = "9",
    year = "1993"
}

@article{Hull:2004in,
    author = "Hull, C. M.",
    title = "{A Geometry for non-geometric string backgrounds}",
    eprint = "hep-th/0406102",
    archivePrefix = "arXiv",
    reportNumber = "IMPERIAL-TP-3-04-13",
    doi = "10.1088/1126-6708/2005/10/065",
    journal = "JHEP",
    volume = "10",
    pages = "065",
    year = "2005"
}

@article{Hull:2006va,
    author = "Hull, C M",
    title = "{Doubled Geometry and T-Folds}",
    eprint = "hep-th/0605149",
    archivePrefix = "arXiv",
    reportNumber = "IMPERIAL-TP-06-CH-02",
    doi = "10.1088/1126-6708/2007/07/080",
    journal = "JHEP",
    volume = "07",
    pages = "080",
    year = "2007"
}

@article{Hull:2009mi,
    author = {Hull, Chris and Zwiebach, Barton},
    journal = {JHEP},
    title = {{Double Field Theory}},
    year = {2009},
    pages = {099},
    volume = {09},
    number = {09},
    archiveprefix = {arXiv},
    doi = {10.1088/1126-6708/2009/09/099},
    eprint = {0904.4664},
    primaryclass = {hep-th},
    reportnumber = {IMPERIAL-TP-2009-CH-02, MIT-CTP-4031},
}

@article{Hull:2009zb,
    author = "Hull, Chris and Zwiebach, Barton",
    title = "{The Gauge algebra of double field theory and Courant brackets}",
    eprint = "0908.1792",
    archivePrefix = "arXiv",
    primaryClass = "hep-th",
    reportNumber = "IMPERIAL-TP-2009-CH-04, MIT-CTP-4054",
    doi = "10.1088/1126-6708/2009/09/090",
    journal = "JHEP",
    volume = "09",
    pages = "090",
    year = "2009"
}

@article{Hohm:2010jy,
    author = "Hohm, Olaf and Hull, Chris and Zwiebach, Barton",
    title = "{Background independent action for double field theory}",
    eprint = "1003.5027",
    archivePrefix = "arXiv",
    primaryClass = "hep-th",
    reportNumber = "IMPERIAL-TP-2010-CH-01, MIT-CTP-4128",
    doi = "10.1007/JHEP07(2010)016",
    journal = "JHEP",
    volume = "07",
    pages = "016",
    year = "2010"
}

@article{Polacek:2013nla,
    author = {Pol\'a\v{c}ek, Martin and Siegel, Warren},
    journal = {JHEP},
    title = {{Natural curvature for manifest T-duality}},
    year = {2014},
    pages = {026},
    volume = {01},
    number = {01},
    archiveprefix = {arXiv},
    doi = {10.1007/JHEP01(2014)026},
    eprint = {1308.6350},
    primaryclass = {hep-th},
}

@article{Butter:2022iza,
    author = {Butter, Daniel and Hassler, Falk and Pope, Christopher N. and
              Zhang, Haoyu},
    journal = {JHEP},
    title = {{Consistent truncations and dualities}},
    year = {2023},
    pages = {007},
    volume = {04},
    number = {04},
    archiveprefix = {arXiv},
    doi = {10.1007/JHEP04(2023)007},
    eprint = {2211.13241},
    primaryclass = {hep-th},
    reportnumber = {MI-HET-788},
}

@article{Geissbuhler:2013uka,
    author = {Geissbuhler, David and Marques, Diego and Nunez, Carmen and Penas,
              Victor},
    journal = {JHEP},
    title = {{Exploring Double Field Theory}},
    year = {2013},
    pages = {101},
    volume = {06},
    number = {06},
    archiveprefix = {arXiv},
    doi = {10.1007/JHEP06(2013)101},
    eprint = {1304.1472},
    primaryclass = {hep-th},
}

@article{Hohm:2010pp,
    author = {Hohm, Olaf and Hull, Chris and Zwiebach, Barton},
    journal = {JHEP},
    title = {{Generalized metric formulation of double field theory}},
    year = {2010},
    pages = {008},
    volume = {08},
    number = {08},
    archiveprefix = {arXiv},
    doi = {10.1007/JHEP08(2010)008},
    eprint = {1006.4823},
    primaryclass = {hep-th},
    reportnumber = {IMPERIAL-TP-2010-CH-03, MIT-CTP-4154},
}

@article{Hitchin:2003cxu,
    author = {Hitchin, Nigel},
    journal = {Quart. J. Math. Oxford Ser.},
    title = {{Generalized Calabi-Yau manifolds}},
    year = {2003},
    pages = {281--308},
    volume = {54},
    number = {54},
    archiveprefix = {arXiv},
    doi = {10.1093/qjmath/54.3.281},
    eprint = {math/0209099},
}

@phdthesis{Gualtieri:2003dx,
    author = {Gualtieri, Marco},
    school = {Oxford U.},
    title = {{Generalized complex geometry}},
    year = {2003},
    archiveprefix = {arXiv},
    eprint = {math/0401221},
}

@article{grutzmann2021weyl,
  title={Weyl quantization of degree 2 symplectic graded manifolds},
  author={Gr{\"u}tzmann, Melchior and Michel, Jean-Philippe and Xu, Ping},
  journal={Journal de Math{\'e}matiques Pures et Appliqu{\'e}es},
  volume={154},
  pages={67--107},
  year={2021},
  publisher={Elsevier}
}

@article{Arvanitakis:2018cyo,
    author = "Arvanitakis, Alex S.",
    title = "{Brane Wess-Zumino terms from AKSZ and exceptional generalised geometry as an $L_\infty$-algebroid}",
    eprint = "1804.07303",
    archivePrefix = "arXiv",
    primaryClass = "hep-th",
    reportNumber = "Imperial-TP-2018-ASA-02, IMPERIAL-TP-2018-ASA-02",
    doi = "10.4310/ATMP.2019.v23.n5.a1",
    journal = "Adv. Theor. Math. Phys.",
    volume = "23",
    number = "5",
    pages = "1159--1213",
    year = "2019"
}

@article{Hassler:2025rag,
    author = "Hassler, Falk and Osten, David and Swash, Alex",
    title = "{Gauged Extended Field Theory and Generalised Cartan Geometry}",
    eprint = "2509.04595",
    archivePrefix = "arXiv",
    primaryClass = "hep-th",
    doi = "10.1103/pzq2-1rhv",
    journal = "Phys. Rev. D",
    volume = "113",
    pages = "066017",
    year = "2026"
}

@article{Hassler:2024hgq,
    author = "Hassler, Falk and Hulik, Ondrej and Osten, David",
    title = "{Current algebra and generalized Cartan geometry}",
    eprint = "2409.00176",
    archivePrefix = "arXiv",
    primaryClass = "hep-th",
    doi = "10.1103/PhysRevD.110.126022",
    journal = "Phys. Rev. D",
    volume = "110",
    number = "12",
    pages = "126022",
    year = "2024"
}

@article{Hassler:2023axp,
    author = "Hassler, Falk and Sakatani, Yuho",
    title = "{Hierarchy of curvatures in exceptional geometry}",
    eprint = "2311.12095",
    archivePrefix = "arXiv",
    primaryClass = "hep-th",
    doi = "10.1103/PhysRevD.109.106002",
    journal = "Phys. Rev. D",
    volume = "109",
    number = "10",
    pages = "106002",
    year = "2024"
}

@article{Osten:2025lsj,
    author = "Osten, David",
    title = "{Generalised Cartan Geometry}",
    eprint = "2605.21809",
    archivePrefix = "arXiv",
    primaryClass = "hep-th",
    journal = "PoS",
    volume = "CORFU2025",
    pages = "373",
    year = "2025"
}

@article{Severa:2017oew,
    author = "{\v{S}}evera, Pavol",
    title = "{Letters to Alan Weinstein about Courant algebroids}",
    eprint = "1707.00265",
    archivePrefix = "arXiv",
    primaryClass = "math.DG",
    month = "7",
    year = "2017"
}

@phdthesis{Roytenberg:1999mny,
    author = "Roytenberg, Dmitry",
    title = "{Courant algebroids, derived brackets and even symplectic supermanifolds}",
    school = "{University of California, Berkeley}",
    year = "1999",
    month = "10",
    eprint = "math/9910078",
    archivePrefix = "arXiv",
    primaryClass = "math.DG"
}

@book{Sharpe:1997,
  author       = {Sharpe, Richard W.},
  title        = {Differential Geometry: Cartan’s Generalization of Klein’s Erlangen Program},
  series       = {Graduate Texts in Mathematics},
  volume       = {166},
  publisher    = {Springer},
  year         = {1997},
}

@Book{		  cap2009parabolic,
  title		= {Parabolic Geometries: Background and general theory},
  author	= {Cap, A. and Slov{\'a}k, J.},
  isbn		= {9780821875353},
  series	= {Mathematical surveys and monographs},
  url		= {https://books.google.be/books?id=G4Ot397nWsQC},
  year		= {2009},
  publisher	= {American Mathematical Society}
}

@article{kosmann2004derived,
  title={Derived brackets},
  author={Kosmann-Schwarzbach, Yvette},
  journal={Letters in Mathematical Physics},
  volume={69},
  number={1},
  pages={61--87},
  year={2004},
  publisher={Springer}
}

@article{roytenberg1998courant,
  title={Courant algebroids and strongly homotopy {Lie} algebras},
  author={Roytenberg, Dmitry and Weinstein, Alan},
  journal={Letters in Mathematical Physics},
  volume={46},
  number={1},
  pages={81--93},
  year={1998},
  publisher={Springer}
}

@article{Cederwall:2021xqi,
    author = "Cederwall, Martin and Palmkvist, Jakob",
    title = "{Teleparallelism in the algebraic approach to extended geometry}",
    eprint = "2112.08403",
    archivePrefix = "arXiv",
    primaryClass = "hep-th",
    doi = "10.1007/JHEP04(2022)164",
    journal = "JHEP",
    volume = "04",
    pages = "164",
    year = "2022"
}

@article{Cederwall:2023xbj,
    author = "Cederwall, Martin and Palmkvist, Jakob",
    title = "{The teleparallel complex}",
    eprint = "2303.15391",
    archivePrefix = "arXiv",
    primaryClass = "hep-th",
    doi = "10.1007/JHEP05(2023)068",
    journal = "JHEP",
    volume = "05",
    pages = "068",
    year = "2023"
}

@article{Cederwall:2025iyq,
    author = "Cederwall, Martin",
    title = "{BV actions for extended geometry}",
    eprint = "2504.20873",
    archivePrefix = "arXiv",
    primaryClass = "hep-th",
    doi = "10.22323/1.490.0333",
    journal = "PoS",
    volume = "CORFU2024",
    pages = "333",
    year = "2025"
}

@inproceedings{Roytenberg:2002nu,
    author = "Roytenberg, Dmitry",
    title = "{On the structure of graded symplectic supermanifolds and Courant algebroids}",
    booktitle = "{Workshop on Quantization, Deformations, and New Homological and Categorical Methods in Mathematical Physics}",
    eprint = "math/0203110",
    archivePrefix = "arXiv",
    month = "3",
    year = "2002"
}

@article{cueca2026lecturenotessymplecticgeometry,
      title={Lecture notes on the symplectic geometry of graded manifolds and higher {Lie} groupoids}, 
      author={Miquel Cueca and Antonio Maglio and Fabricio Valencia},
      year={2026},
      eprint={2510.09448},
      archivePrefix={arXiv},
      primaryClass={math.SG},
      url={https://arxiv.org/abs/2510.09448}, 
}

@article{Mehta_2013,
   title={{Differential Graded Contact Geometry and Jacobi Structures}},
   volume={103},
   ISSN={1573-0530},
   url={http://dx.doi.org/10.1007/s11005-013-0609-6},
   DOI={10.1007/s11005-013-0609-6},
   number={7},
   journal={Letters in Mathematical Physics},
   publisher={Springer Science and Business Media LLC},
   author={Mehta, Rajan Amit},
   year={2013},
   month=Jan, pages={729–741} }

@article{Kotov:2007nr,
    author = "Kotov, Alexei and Strobl, Thomas",
    title = "{Characteristic classes associated to Q-bundles}",
    eprint = "0711.4106",
    archivePrefix = "arXiv",
    primaryClass = "math.DG",
    doi = "10.1142/S0219887815500061",
    journal = "Int. J. Geom. Meth. Mod. Phys.",
    volume = "12",
    number = "01",
    pages = "1550006",
    year = "2014"
}

@article{keller2015deformation,
  title={Deformation theory of {Courant algebroids via the Rothstein} algebra},
  author={Keller, Frank and Waldmann, Stefan},
  journal={Journal of Pure and Applied Algebra},
  volume={219},
  number={8},
  pages={3391--3426},
  year={2015},
  publisher={Elsevier}
}

@article{Deser:2014mxa,
    author = "Deser, Andreas and Stasheff, Jim",
    title = "{Even symplectic supermanifolds and double field theory}",
    eprint = "1406.3601",
    archivePrefix = "arXiv",
    primaryClass = "math-ph",
    reportNumber = "ITP-UH-07-14",
    doi = "10.1007/s00220-015-2443-4",
    journal = "Commun. Math. Phys.",
    volume = "339",
    number = "3",
    pages = "1003--1020",
    year = "2015"
}

@article{Deser_2018,
   title={Extended {Riemannian Geometry I: Local Double Field Theory}},
   volume={19},
   ISSN={1424-0661},
   url={http://dx.doi.org/10.1007/s00023-018-0694-2},
   DOI={10.1007/s00023-018-0694-2},
   number={8},
   journal={Annales Henri Poincaré},
   publisher={Springer Science and Business Media LLC},
   author={Deser, Andreas and Sämann, Christian},
   year={2018}, pages={2297–2346},
   eprint = "1611.02772",
   archivePrefix = "arXiv",
   primaryClass = "hep-th"
}

@article{Deser:2017fko,
    author = {Deser, Andreas and Heller, Marc Andre and S{\"a}mann, Christian},
    title = "{Extended Riemannian Geometry II: Local Heterotic Double Field Theory}",
    eprint = "1711.03308",
    archivePrefix = "arXiv",
    primaryClass = "hep-th",
    reportNumber = "EMPG-17-18",
    doi = "10.1007/JHEP04(2018)106",
    journal = "JHEP",
    volume = "04",
    pages = "106",
    year = "2018"
}

@article{Deser:2018flj,
    author = {Deser, Andreas and S{\"a}mann, Christian},
    title = "{Extended Riemannian Geometry III: Global Double Field Theory with Nilmanifolds}",
    eprint = "1812.00026",
    archivePrefix = "arXiv",
    primaryClass = "hep-th",
    reportNumber = "EMPG-18-24",
    doi = "10.1007/JHEP05(2019)209",
    journal = "JHEP",
    volume = "05",
    pages = "209",
    year = "2019"
}

@article{BAYEN197861,
title = {{Deformation theory and quantization. I. Deformations of symplectic structures}},
journal = {Annals of Physics},
volume = {111},
number = {1},
pages = {61-110},
year = {1978},
issn = {0003-4916},
doi = {https://doi.org/10.1016/0003-4916(78)90224-5},
url = {https://www.sciencedirect.com/science/article/pii/0003491678902245},
author = {F Bayen and M Flato and C Fronsdal and A Lichnerowicz and D Sternheimer}
}

@article{Bordemann:1999ca,
    author = "Bordemann, M.",
    editor = "Dito, Giuseppe and Sternheimer, Daniel",
    title = "{The deformation quantization of certain super-Poisson brackets and BRST cohomology}",
    doi = "10.1007/978-94-015-1276-3_4",
    journal = "Math. Phys. Stud.",
    volume = "21-22",
    pages = "45--68",
    year = "2000"
}

@article{Hirshfeld:2002ki,
    author = "Hirshfeld, A. C. and Henselder, P.",
    title = "{Deformation quantization for systems with fermions}",
    doi = "10.1006/aphy.2002.6302",
    journal = "Annals Phys.",
    volume = "302",
    pages = "59--77",
    year = "2002"
}

@article{henselder2005star,
  title={Star products and geometric algebra},
  author={Henselder, Peter and Hirshfeld, Allen C and Spernat, Thomas},
  journal={Annals of Physics},
  volume={317},
  number={1},
  pages={107--129},
  year={2005},
  publisher={Elsevier}
}

@article{Bergshoeff:2019sfy,
    author = "Bergshoeff, Eric and Kleinschmidt, Axel and Musaev, Edvard T. and Riccioni, Fabio",
    title = "{The different faces of branes in Double Field Theory}",
    eprint = "1903.05601",
    archivePrefix = "arXiv",
    primaryClass = "hep-th",
    doi = "10.1007/JHEP09(2019)110",
    journal = "JHEP",
    volume = "09",
    pages = "110",
    year = "2019"
}

@article{Bergshoeff:2016ncb,
    author = "Bergshoeff, Eric A. and Hohm, Olaf and Penas, Victor A. and Riccioni, Fabio",
    title = "{Dual Double Field Theory}",
    eprint = "1603.07380",
    archivePrefix = "arXiv",
    primaryClass = "hep-th",
    doi = "10.1007/JHEP06(2016)026",
    journal = "JHEP",
    volume = "06",
    pages = "026",
    year = "2016"
}

@article{Duff:1989tf,
    author = "Duff, M. J.",
    editor = "Pati, Jogesh C. and Randjbar- Daemi, S. and Sezgin, E. and Shafi, Q.",
    title = "{Duality Rotations in String Theory}",
    reportNumber = "CTP-TAMU-53-89",
    doi = "10.1016/0550-3213(90)90520-N",
    journal = "Nucl. Phys. B",
    volume = "335",
    pages = "610",
    year = "1990"
}

@article{Tseytlin:1990nb,
    author = "Tseytlin, Arkady A.",
    title = "{Duality Symmetric Formulation of String World Sheet Dynamics}",
    reportNumber = "KCL-TP-1990-2",
    doi = "10.1016/0370-2693(90)91454-J",
    journal = "Phys. Lett. B",
    volume = "242",
    pages = "163--174",
    year = "1990"
}

@article{Tseytlin:1990va,
    author = "Tseytlin, Arkady A.",
    title = "{Duality symmetric closed string theory and interacting chiral scalars}",
    reportNumber = "KCL-TP-1990-3",
    doi = "10.1016/0550-3213(91)90266-Z",
    journal = "Nucl. Phys. B",
    volume = "350",
    pages = "395--440",
    year = "1991"
}

@article{Aldazabal:2013sca,
    author = "Aldazabal, Gerardo and Marques, Diego and Nunez, Carmen",
    title = "{Double Field Theory: A Pedagogical Review}",
    eprint = "1305.1907",
    archivePrefix = "arXiv",
    primaryClass = "hep-th",
    doi = "10.1088/0264-9381/30/16/163001",
    journal = "Class. Quant. Grav.",
    volume = "30",
    pages = "163001",
    year = "2013"
}

@article{Berman:2013eva,
    author = "Berman, David S. and Thompson, Daniel C.",
    title = "{Duality Symmetric String and M-Theory}",
    eprint = "1306.2643",
    archivePrefix = "arXiv",
    primaryClass = "hep-th",
    doi = "10.1016/j.physrep.2014.11.007",
    journal = "Phys. Rept.",
    volume = "566",
    pages = "1--60",
    year = "2014"
}

@article{Hohm:2013bwa,
    author = {Hohm, Olaf and L{\"u}st, Dieter and Zwiebach, Barton},
    title = "{The Spacetime of Double Field Theory: Review, Remarks, and Outlook}",
    eprint = "1309.2977",
    archivePrefix = "arXiv",
    primaryClass = "hep-th",
    reportNumber = "MIT-CTP-4494, LMU-ASC-59-13, MPP-2013-241",
    doi = "10.1002/prop.201300024",
    journal = "Fortsch. Phys.",
    volume = "61",
    pages = "926--966",
    year = "2013"
}

@article{Giveon:1994fu,
    author = "Giveon, Amit and Porrati, Massimo and Rabinovici, Eliezer",
    title = "{Target space duality in string theory}",
    eprint = "hep-th/9401139",
    archivePrefix = "arXiv",
    reportNumber = "RI-1-94, NYU-TH-94-01-01",
    doi = "10.1016/0370-1573(94)90070-1",
    journal = "Phys. Rept.",
    volume = "244",
    pages = "77--202",
    year = "1994"
}

@article{Berman:2012vc,
    author = "Berman, David S. and Cederwall, Martin and Kleinschmidt, Axel and Thompson, Daniel C.",
    title = "{The gauge structure of generalised diffeomorphisms}",
    eprint = "1208.5884",
    archivePrefix = "arXiv",
    primaryClass = "hep-th",
    reportNumber = "QMUL-PH-12-14, AEI-2012-085",
    doi = "10.1007/JHEP01(2013)064",
    journal = "JHEP",
    volume = "01",
    pages = "064",
    year = "2013"
}

@article{Batalin:1981jr,
    author = "Batalin, I. A. and Vilkovisky, G. A.",
    title = "{Gauge Algebra and Quantization}",
    doi = "10.1016/0370-2693(81)90205-7",
    journal = "Phys. Lett. B",
    volume = "102",
    pages = "27--31",
    year = "1981"
}

@article{Batalin:1983ggl,
    author = "Batalin, I. A. and Vilkovisky, G. A.",
    title = "{Quantization of Gauge Theories with Linearly Dependent Generators}",
    doi = "10.1103/PhysRevD.28.2567",
    journal = "Phys. Rev. D",
    volume = "28",
    pages = "2567--2582",
    year = "1983",
    note = "[Erratum: Phys.Rev.D 30, 508 (1984)]"
}

@article{Jurco:2018sby,
    author = {Jur{\v{c}}o, Branislav and Raspollini, Lorenzo and S{\"a}mann, Christian and Wolf, Martin},
    title = "{$L_\infty$-Algebras of Classical Field Theories and the Batalin-Vilkovisky Formalism}",
    eprint = "1809.09899",
    archivePrefix = "arXiv",
    primaryClass = "hep-th",
    reportNumber = "EMPG-18-19, DMUS-MP-18/05",
    doi = "10.1002/prop.201900025",
    journal = "Fortsch. Phys.",
    volume = "67",
    number = "7",
    pages = "1900025",
    year = "2019"
}

@article{Schwarz:1992nx,
    author = "Schwarz, Albert S.",
    title = "{Geometry of Batalin-Vilkovisky quantization}",
    eprint = "hep-th/9205088",
    archivePrefix = "arXiv",
    reportNumber = "PRINT-92-0192 (UC,DAVIS)",
    doi = "10.1007/BF02097392",
    journal = "Commun. Math. Phys.",
    volume = "155",
    pages = "249--260",
    year = "1993"
}

@article{Alexandrov:1995kv,
    author = "Alexandrov, M. and Schwarz, A. and Zaboronsky, O. and Kontsevich, M.",
    title = "{The Geometry of the master equation and topological quantum field theory}",
    eprint = "hep-th/9502010",
    archivePrefix = "arXiv",
    reportNumber = "UCD-94-01, UCD{\&}B-94-01",
    doi = "10.1142/S0217751X97001031",
    journal = "Int. J. Mod. Phys. A",
    volume = "12",
    pages = "1405--1429",
    year = "1997"
}

@article{Liu:1995lsa,
    author = "Liu, Zhang-Ju and Weinstein, Alan and Xu, Ping",
    title = "{Manin Triples for Lie Bialgebroids}",
    eprint = "dg-ga/9508013",
    archivePrefix = "arXiv",
    journal = "J. Diff. Geom.",
    volume = "45",
    number = "3",
    pages = "547--574",
    year = "1997"
}

@article{Palmkvist:2013vya,
    author = "Palmkvist, Jakob",
    title = "{The tensor hierarchy algebra}",
    eprint = "1305.0018",
    archivePrefix = "arXiv",
    primaryClass = "hep-th",
    doi = "10.1063/1.4858335",
    journal = "J. Math. Phys.",
    volume = "55",
    pages = "011701",
    year = "2014"
}

@article{Greitz:2013pua,
    author = "Greitz, Jesper and Howe, Paul and Palmkvist, Jakob",
    title = "{The tensor hierarchy simplified}",
    eprint = "1308.4972",
    archivePrefix = "arXiv",
    primaryClass = "hep-th",
    reportNumber = "KCL-MTH-13-08, NORDITA-2013-062, IHES-P-13-27",
    doi = "10.1088/0264-9381/31/8/087001",
    journal = "Class. Quant. Grav.",
    volume = "31",
    pages = "087001",
    year = "2014"
}

@article{Palmkvist:2015dea,
    author = "Palmkvist, Jakob",
    title = "{Exceptional geometry and Borcherds superalgebras}",
    eprint = "1507.08828",
    archivePrefix = "arXiv",
    primaryClass = "hep-th",
    reportNumber = "MI-TH-1520",
    doi = "10.1007/JHEP11(2015)032",
    journal = "JHEP",
    volume = "11",
    pages = "032",
    year = "2015"
}

@article{Cederwall:2017fjm,
    author = "Cederwall, Martin and Palmkvist, Jakob",
    title = "{Extended geometries}",
    eprint = "1711.07694",
    archivePrefix = "arXiv",
    primaryClass = "hep-th",
    doi = "10.1007/JHEP02(2018)071",
    journal = "JHEP",
    volume = "02",
    pages = "071",
    year = "2018"
}

@article{Cederwall:2018aab,
    author = "Cederwall, Martin and Palmkvist, Jakob",
    title = "{$L_{\infty }$ Algebras for Extended Geometry from Borcherds Superalgebras}",
    eprint = "1804.04377",
    archivePrefix = "arXiv",
    primaryClass = "hep-th",
    doi = "10.1007/s00220-019-03451-2",
    journal = "Commun. Math. Phys.",
    volume = "369",
    number = "2",
    pages = "721--760",
    year = "2019"
}

@article{Cederwall:2019qnw,
    author = "Cederwall, Martin and Palmkvist, Jakob",
    title = "{Tensor hierarchy algebras and extended geometry. Part I. Construction of the algebra}",
    eprint = "1908.08695",
    archivePrefix = "arXiv",
    primaryClass = "hep-th",
    doi = "10.1007/JHEP02(2020)144",
    journal = "JHEP",
    volume = "02",
    pages = "144",
    year = "2020"
}

@article{Cederwall:2019bai,
    author = "Cederwall, Martin and Palmkvist, Jakob",
    title = "{Tensor hierarchy algebras and extended geometry. Part II. Gauge structure and dynamics}",
    eprint = "1908.08696",
    archivePrefix = "arXiv",
    primaryClass = "hep-th",
    doi = "10.1007/JHEP02(2020)145",
    journal = "JHEP",
    volume = "02",
    pages = "145",
    year = "2020"
}

@article{Stasheff:1997iz,
    author = "Stasheff, Jim",
    editor = "Henneaux, Marc and Krasil'shchik, Joseph and Vinogradov, Alexandre",
    title = "{The (Secret?) homological algebra of the Batalin-Vilkovisky approach}",
    eprint = "hep-th/9712157",
    archivePrefix = "arXiv",
    doi = "10.1090/conm/219/03076",
    journal = "Contemp. Math.",
    volume = "219",
    pages = "195--210",
    year = "1998"
}

@article{Hohm:2017pnh,
    author = "Hohm, Olaf and Zwiebach, Barton",
    title = "{$L_{\infty}$ Algebras and Field Theory}",
    eprint = "1701.08824",
    archivePrefix = "arXiv",
    primaryClass = "hep-th",
    reportNumber = "MIT-CTP-4875",
    doi = "10.1002/prop.201700014",
    journal = "Fortsch. Phys.",
    volume = "65",
    number = "3-4",
    pages = "1700014",
    year = "2017"
}

@article{Heller:2016abk,
    author = "Heller, Marc Andre and Ikeda, Noriaki and Watamura, Satoshi",
    title = "{Unified picture of non-geometric fluxes and T-duality in double field theory via graded symplectic manifolds}",
    eprint = "1611.08346",
    archivePrefix = "arXiv",
    primaryClass = "hep-th",
    reportNumber = "TU-1036",
    doi = "10.1007/JHEP02(2017)078",
    journal = "JHEP",
    volume = "02",
    pages = "078",
    year = "2017"
}

@article{Carow-Watamura:2018iau,
    author = "Carow-Watamura, Ursula and Ikeda, Noriaki and Kaneko, Tomokazu and Watamura, Satoshi",
    title = "{DFT in supermanifold formulation and group manifold as background geometry}",
    eprint = "1812.03464",
    archivePrefix = "arXiv",
    primaryClass = "hep-th",
    reportNumber = "TU-1078",
    doi = "10.1007/JHEP04(2019)002",
    journal = "JHEP",
    volume = "04",
    pages = "002",
    year = "2019"
}

@unpublished{Alekseev2001,
  author = {Alekseev, A. and Xu, P.},
  title = "{Derived brackets and Courant algebroids}",
  year = {2001},
  note = {Unpublished manuscript}
}

@article{Carow-Watamura:2020xij,
    author = "Carow-Watamura, Ursula and Miura, Kohei and Watamura, Satoshi and Yano, Taro",
    title = "{Metric algebroid and Dirac generating operator in Double Field Theory}",
    eprint = "2005.04658",
    archivePrefix = "arXiv",
    primaryClass = "hep-th",
    reportNumber = "preprint TU-1101",
    doi = "10.1007/JHEP10(2020)192",
    journal = "JHEP",
    volume = "10",
    pages = "192",
    year = "2020"
}

@article{Carow-Watamura:2022ten,
    author = "Carow-Watamura, Ursula and Miura, Kohei and Watamura, Satoshi",
    title = "{Metric Algebroid and Poisson-Lie T-duality in DFT}",
    eprint = "2207.14725",
    archivePrefix = "arXiv",
    primaryClass = "hep-th",
    reportNumber = "TU-1161",
    doi = "10.1007/s00220-023-04765-y",
    journal = "Commun. Math. Phys.",
    volume = "402",
    number = "2",
    pages = "1879--1930",
    year = "2023"
}

@article{Kontsevich:1997vb,
    author = "Kontsevich, Maxim",
    title = "{Deformation quantization of Poisson manifolds. 1.}",
    eprint = "q-alg/9709040",
    archivePrefix = "arXiv",
    doi = "10.1023/B:MATH.0000027508.00421.bf",
    journal = "Lett. Math. Phys.",
    volume = "66",
    pages = "157--216",
    year = "2003"
}

@article{Fedosov:1994zz,
    author = "Fedosov, Boris v.",
    title = "{A Simple geometrical construction of deformation quantization}",
    journal = "J. Diff. Geom.",
    volume = "40",
    number = "2",
    pages = "213--238",
    year = "1994"
}

@ARTICLE{1983LMaPh...7..487D,
       author = {{de Wilde}, Marc and {Lecomte}, Pierre B.~A.},
        title = "{Existence of star-products and of formal deformations of the Poisson Lie algebra of arbitrary symplectic manifolds}",
      journal = {Letters in Mathematical Physics},
         year = 1983,
        month = nov,
       volume = {7},
       number = {6},
        pages = {487-496},
          doi = {10.1007/BF00402248},
       adsurl = {https://ui.adsabs.harvard.edu/abs/1983LMaPh...7..487D}
}

@article{Osten:2023iwc,
    author = "Osten, David",
    title = "{On exceptional QP-manifolds}",
    eprint = "2306.11093",
    archivePrefix = "arXiv",
    primaryClass = "hep-th",
    doi = "10.1007/JHEP01(2024)028",
    journal = "JHEP",
    volume = "01",
    pages = "028",
    year = "2024"
}

@article{grabowski2013graded,
  title={Graded contact manifolds and contact {Courant} algebroids},
  author={Grabowski, Janusz},
  journal={Journal of Geometry and Physics},
  volume={68},
  pages={27--58},
  year={2013},
  publisher={Elsevier}
}

@article{contreras2026graded,
  title={Graded contact geometry and the {AKSZ} formalism},
  author={Contreras, Ivan and Alba, Nicolas Martinez and Mehta, Rajan Amit},
  journal={Differential Geometry and its Applications},
  volume={103},
  pages={102359},
  year={2026},
  publisher={Elsevier}
}

@article{Barnich:2000zw,
    author = "Barnich, Glenn and Brandt, Friedemann and Henneaux, Marc",
    title = "{Local BRST cohomology in gauge theories}",
    eprint = "hep-th/0002245",
    archivePrefix = "arXiv",
    reportNumber = "ITP-UH-03-00, ULB-TH-00-05",
    doi = "10.1016/S0370-1573(00)00049-1",
    journal = "Phys. Rept.",
    volume = "338",
    pages = "439--569",
    year = "2000"
}

@article{Hohm:2013nja,
    author = "Hohm, Olaf and Samtleben, Henning",
    title = "{Gauge theory of Kaluza-Klein and winding modes}",
    eprint = "1307.0039",
    archivePrefix = "arXiv",
    primaryClass = "hep-th",
    reportNumber = "LMU-ASC-44-13",
    doi = "10.1103/PhysRevD.88.085005",
    journal = "Phys. Rev. D",
    volume = "88",
    pages = "085005",
    year = "2013"
}

@article{Tyutin:2001iz,
    author = "Tyutin, I. V.",
    title = "{The General form of the star product on the Grassman algebra}",
    eprint = "hep-th/0101046",
    archivePrefix = "arXiv",
    reportNumber = "FIAN-TD-01-01",
    doi = "10.1023/A:1010445502845",
    journal = "Theor. Math. Phys.",
    volume = "127",
    pages = "619--631",
    year = "2001"
}

@article{Hassler:2017yza,
    author = "Hassler, Falk",
    title = "{Poisson-Lie T-duality in Double Field Theory}",
    eprint = "1707.08624",
    archivePrefix = "arXiv",
    primaryClass = "hep-th",
    doi = "10.1016/j.physletb.2020.135455",
    journal = "Phys. Lett. B",
    volume = "807",
    pages = "135455",
    year = "2020"
}

@article{Hohm:2011zr,
    author = "Hohm, Olaf and Kwak, Seung Ki and Zwiebach, Barton",
    title = "{Unification of Type II Strings and T-duality}",
    eprint = "1106.5452",
    archivePrefix = "arXiv",
    primaryClass = "hep-th",
    reportNumber = "MIT-CTP-4277",
    doi = "10.1103/PhysRevLett.107.171603",
    journal = "Phys. Rev. Lett.",
    volume = "107",
    pages = "171603",
    year = "2011"
}

@article{Hohm:2011dv,
    author = "Hohm, Olaf and Kwak, Seung Ki and Zwiebach, Barton",
    title = "{Double Field Theory of Type II Strings}",
    eprint = "1107.0008",
    archivePrefix = "arXiv",
    primaryClass = "hep-th",
    reportNumber = "MIT-CTP-4278",
    doi = "10.1007/JHEP09(2011)013",
    journal = "JHEP",
    volume = "09",
    pages = "013",
    year = "2011"
}

@article{Jeon:2011vx,
    author = "Jeon, Imtak and Lee, Kanghoon and Park, Jeong-Hyuck",
    title = "{Incorporation of fermions into double field theory}",
    eprint = "1109.2035",
    archivePrefix = "arXiv",
    primaryClass = "hep-th",
    doi = "10.1007/JHEP11(2011)025",
    journal = "JHEP",
    volume = "11",
    pages = "025",
    year = "2011"
}

@article{Jeon:2011sq,
    author = "Jeon, Imtak and Lee, Kanghoon and Park, Jeong-Hyuck",
    title = "{Supersymmetric Double Field Theory: Stringy Reformulation of Supergravity}",
    eprint = "1112.0069",
    archivePrefix = "arXiv",
    primaryClass = "hep-th",
    reportNumber = "CERN-PH-TH-2011-278",
    doi = "10.1103/PhysRevD.85.081501",
    journal = "Phys. Rev. D",
    volume = "85",
    pages = "081501",
    year = "2012",
    note = "[Erratum: Phys.Rev.D 86, 089903 (2012)]"
}

@article{Jeon:2012kd,
    author = "Jeon, Imtak and Lee, Kanghoon and Park, Jeong-Hyuck",
    title = "{Ramond-Ramond Cohomology and O(D,D) T-duality}",
    eprint = "1206.3478",
    archivePrefix = "arXiv",
    primaryClass = "hep-th",
    doi = "10.1007/JHEP09(2012)079",
    journal = "JHEP",
    volume = "09",
    pages = "079",
    year = "2012"
}

@inproceedings{Ikeda:2012pv,
    author = "Ikeda, Noriaki",
    title = "{Lectures on AKSZ Sigma Models for Physicists}",
    booktitle = "{Workshop on Strings, Membranes and Topological Field Theory}",
    eprint = "1204.3714",
    archivePrefix = "arXiv",
    primaryClass = "hep-th",
    doi = "10.1142/9789813144613_0003",
    publisher = "WSPC",
    pages = "79--169",
    year = "2017"
}

@InProceedings{10.1007/978-3-031-89857-0_10,
author="Ikeda, Noriaki",
editor="Kielanowski, Piotr
and Dobrogowska, Alina
and Fern{\'a}ndez, David
and Goli{\'{n}}ski, Tomasz",
title="{Q-Manifolds and Sigma Models}",
booktitle="Geometric Methods in Physics XLI",
year="2025",
publisher="Springer Nature Switzerland",
address="Cham",
pages="119--140",
isbn="978-3-031-89857-0"
}

@article{CuecaMehta2021,
  author  = {Cueca, Miquel and Mehta, Rajan Amit},
  title   = {{Courant Cohomology, Cartan Calculus, Connections, Curvature, Characteristic Classes}},
  journal = {Communications in Mathematical Physics},
  volume  = {381},
  number  = {3},
  pages   = {1091--1113},
  year    = {2021},
  month   = feb,
  doi     = {10.1007/s00220-020-03894-y}
}

@article{Cattaneo:2010re,
    author = "Cattaneo, Alberto S. and Schaetz, Florian",
    title = "{Introduction to supergeometry}",
    eprint = "1011.3401",
    archivePrefix = "arXiv",
    primaryClass = "math-ph",
    doi = "10.1142/S0129055X11004400",
    journal = "Rev. Math. Phys.",
    volume = "23",
    pages = "669--690",
    year = "2011"
}

@article{GinotGrutzmann2009,
  author = {Ginot, Gr{\'e}gory and Gr{\"u}tzmann, Melchior},
  title = {Cohomology of {Courant} algebroids with split base},
  journal = {Journal of Symplectic Geometry},
  volume = {7},
  number = {3},
  pages = {311--335},
  year = {2009},
  month = {September}
}

@article{voronov2005higher,
  title={Higher derived brackets and homotopy algebras},
  author={Voronov, Theodore},
  journal={Journal of pure and applied algebra},
  volume={202},
  number={1-3},
  pages={133--153},
  year={2005},
  publisher={Elsevier}
}

@article{getzler2010higher,
  title={Higher derived brackets},
  author={Getzler, Ezra},
  journal={arXiv preprint arXiv:1010.5859},
  year={2010}
}

@article{fiorenza2007structures,
  title="{$L_\infty$ structures on mapping cones}",
  author={Fiorenza, Domenico and Manetti, Marco},
  journal={Algebra \& Number Theory},
  volume={1},
  number={3},
  pages={301--330},
  year={2007},
  publisher={Mathematical Sciences Publishers}
}

@article{Garcia-Fernandez:2016ofz,
    author = "Garcia-Fernandez, Mario",
    title = "{Ricci flow, Killing spinors, and T-duality in generalized geometry}",
    eprint = "1611.08926",
    archivePrefix = "arXiv",
    primaryClass = "math.DG",
    doi = "10.1016/j.aim.2019.04.038",
    journal = "Adv. Math.",
    volume = "350",
    pages = "1059--1108",
    year = "2019"
}

@article{Boffo:2019zus,
    author = "Boffo, Eugenia and Schupp, Peter",
    title = "{Deformed graded Poisson structures, Generalized Geometry and Supergravity}",
    eprint = "1903.09112",
    archivePrefix = "arXiv",
    primaryClass = "hep-th",
    doi = "10.1007/JHEP01(2020)007",
    journal = "JHEP",
    volume = "01",
    pages = "007",
    year = "2020"
}

@article{Boffo:2021srg,
    author = "Boffo, Eugenia and Schupp, Peter",
    title = "{A gravitational action with stringy Q and R fluxes via deformed differential graded Poisson algebras}",
    eprint = "2106.09601",
    archivePrefix = "arXiv",
    primaryClass = "hep-th",
    doi = "10.1007/JHEP12(2021)143",
    journal = "JHEP",
    volume = "12",
    pages = "143",
    year = "2021"
}

@article{Severa:2018pag,
    author = "{\v{S}}evera, Pavol and Valach, Fridrich",
    title = "{Courant Algebroids, Poisson{\textendash}Lie T-Duality, and Type II Supergravities}",
    eprint = "1810.07763",
    archivePrefix = "arXiv",
    primaryClass = "math.DG",
    doi = "10.1007/s00220-020-03736-x",
    journal = "Commun. Math. Phys.",
    volume = "375",
    number = "1",
    pages = "307--344",
    year = "2020"
}

@article{Ritter:2015ffa,
    author = "Ritter, Patricia and Saemann, Christian",
    title = "{Automorphisms of Strong Homotopy Lie Algebras of Local Observables}",
    eprint = "1507.00972",
    archivePrefix = "arXiv",
    primaryClass = "hep-th",
    reportNumber = "DIFA-2015, EMPG-15-08",
    month = "7",
    year = "2015"
}

@article{Lyakhovich:2004kr,
    author = "Lyakhovich, S. L. and Sharapov, A. A.",
    title = "{Characteristic classes of gauge systems}",
    eprint = "hep-th/0407113",
    archivePrefix = "arXiv",
    doi = "10.1016/j.nuclphysb.2004.10.001",
    journal = "Nucl. Phys. B",
    volume = "703",
    pages = "419--453",
    year = "2004"
}

@article{Lyakhovich:2009qq,
    author = "Lyakhovich, S. L. and Mosman, E. A. and Sharapov, A. A.",
    title = "{Characteristic classes of Q-manifolds: classification and applications}",
    eprint = "0906.0466",
    archivePrefix = "arXiv",
    primaryClass = "math-ph",
    doi = "10.1016/j.geomphys.2010.01.008",
    journal = "J. Geom. Phys.",
    volume = "60",
    pages = "729--759",
    year = "2010"
}

@article{Hohm:2011si,
    author = "Hohm, Olaf and Zwiebach, Barton",
    title = "{On the Riemann Tensor in Double Field Theory}",
    eprint = "1112.5296",
    archivePrefix = "arXiv",
    primaryClass = "hep-th",
    reportNumber = "MIT-CTP-4331, LMU-ASC-75-11",
    doi = "10.1007/JHEP05(2012)126",
    journal = "JHEP",
    volume = "05",
    pages = "126",
    year = "2012"
}

@article{Coimbra:2011nw,
    author = "Coimbra, Andre and Strickland-Constable, Charles and Waldram, Daniel",
    title = "{Supergravity as Generalised Geometry I: Type II Theories}",
    eprint = "1107.1733",
    archivePrefix = "arXiv",
    primaryClass = "hep-th",
    reportNumber = "IMPERIAL-TP-11-DW-01",
    doi = "10.1007/JHEP11(2011)091",
    journal = "JHEP",
    volume = "11",
    pages = "091",
    year = "2011"
}

@incollection{grabowski2006courant,
  author    = {Grabowski, Janusz},
  title     = "{Courant-Nijenhuis Tensors and Generalized Geometries}",
  booktitle = {Groups, Geometry and Physics},
  series    = {Monogr. Real Acad. Ci. Exact. F\'is.-Qu\'im. Nat. Zaragoza},
  volume    = {29},
  pages     = {101--112},
  publisher = {Acad. Cienc. Exact. F\'is. Qu\'im. Nat. Zaragoza},
  address   = {Zaragoza},
  year      = {2006},
  eprint    = {math/0601761},
  archivePrefix = {arXiv}
}

@article{Bering:2006eb,
    author = "Bering, K.",
    title = "{On non-commutative Batalin-Vilkovisky algebras, strongly homotopy Lie algebras and the Courant bracket}",
    eprint = "hep-th/0603116",
    archivePrefix = "arXiv",
    doi = "10.1007/s00220-007-0278-3",
    journal = "Commun. Math. Phys.",
    volume = "274",
    pages = "297--341",
    year = "2007"
}

@article{Osten:2024mjt,
    author = "Osten, David",
    title = "{On the universal exceptional structure of world-volume theories in string and M-theory}",
    eprint = "2402.10269",
    archivePrefix = "arXiv",
    primaryClass = "hep-th",
    doi = "10.1016/j.physletb.2024.138814",
    journal = "Phys. Lett. B",
    volume = "855",
    pages = "138814",
    year = "2024"
}

@article{Alekseev:2004np,
    author = "Alekseev, Anton and Strobl, Thomas",
    title = "{Current algebras and differential geometry}",
    eprint = "hep-th/0410183",
    archivePrefix = "arXiv",
    reportNumber = "FSU-TPI-07-04",
    doi = "10.1088/1126-6708/2005/03/035",
    journal = "JHEP",
    volume = "03",
    pages = "035",
    year = "2005"
}

@article{Arvanitakis:2021wkt,
    author = "Arvanitakis, Alex S.",
    title = "{Brane current algebras and generalised geometry from QP manifolds. Or, {\textquotedblleft}when they go high, we go low{\textquotedblright}}",
    eprint = "2103.08608",
    archivePrefix = "arXiv",
    primaryClass = "hep-th",
    doi = "10.1007/JHEP11(2021)114",
    journal = "JHEP",
    volume = "11",
    pages = "114",
    year = "2021"
}

@article{Ikeda:2011ax,
    author = "Ikeda, Noriaki and Koizumi, Kozo",
    title = "{Current Algebras and QP Manifolds}",
    eprint = "1108.0473",
    archivePrefix = "arXiv",
    primaryClass = "hep-th",
    reportNumber = "MISC-2011-14",
    doi = "10.1142/S0219887813500242",
    journal = "Int. J. Geom. Meth. Mod. Phys.",
    volume = "10",
    pages = "1350024",
    year = "2013"
}

@article{Lavau:2019oja,
    author = "Lavau, Sylvain and Palmkvist, Jakob",
    title = "{Infinity-enhancing of Leibniz algebras}",
    eprint = "1907.05752",
    archivePrefix = "arXiv",
    primaryClass = "hep-th",
    doi = "10.1007/s11005-020-01324-7",
    journal = "Lett. Math. Phys.",
    volume = "110",
    number = "11",
    pages = "3121--3152",
    year = "2020"
}

@article{Hohm:2013pua,
    author = "Hohm, Olaf and Samtleben, Henning",
    title = "{Exceptional Form of D=11 Supergravity}",
    eprint = "1308.1673",
    archivePrefix = "arXiv",
    primaryClass = "hep-th",
    doi = "10.1103/PhysRevLett.111.231601",
    journal = "Phys. Rev. Lett.",
    volume = "111",
    pages = "231601",
    year = "2013"
}

@article{Hohm:2013vpa,
    author = "Hohm, Olaf and Samtleben, Henning",
    title = "{Exceptional Field Theory I: $E_{6(6)}$ covariant Form of M-Theory and Type IIB}",
    eprint = "1312.0614",
    archivePrefix = "arXiv",
    primaryClass = "hep-th",
    reportNumber = "MIT-CTP-4519",
    doi = "10.1103/PhysRevD.89.066016",
    journal = "Phys. Rev. D",
    volume = "89",
    number = "6",
    pages = "066016",
    year = "2014"
}

@article{Hohm:2013uia,
    author = "Hohm, Olaf and Samtleben, Henning",
    title = "{Exceptional field theory. II. E$_{7(7)}$}",
    eprint = "1312.4542",
    archivePrefix = "arXiv",
    primaryClass = "hep-th",
    reportNumber = "MIT-CTP-4522",
    doi = "10.1103/PhysRevD.89.066017",
    journal = "Phys. Rev. D",
    volume = "89",
    pages = "066017",
    year = "2014"
}

@article{Hohm:2014fxa,
    author = "Hohm, Olaf and Samtleben, Henning",
    title = "{Exceptional field theory. III. E$_{8(8)}$}",
    eprint = "1406.3348",
    archivePrefix = "arXiv",
    primaryClass = "hep-th",
    reportNumber = "MIT-CTP-4557",
    doi = "10.1103/PhysRevD.90.066002",
    journal = "Phys. Rev. D",
    volume = "90",
    pages = "066002",
    year = "2014"
}

@article{Ikeda:2002wh,
    author = "Ikeda, Noriaki",
    title = "{Chern-Simons gauge theory coupled with BF theory}",
    eprint = "hep-th/0203043",
    archivePrefix = "arXiv",
    doi = "10.1142/S0217751X03015155",
    journal = "Int. J. Mod. Phys. A",
    volume = "18",
    pages = "2689--2702",
    year = "2003"
}

@article{Roytenberg:2006qz,
    author = "Roytenberg, Dmitry",
    title = "{AKSZ-BV Formalism and Courant Algebroid-induced Topological Field Theories}",
    eprint = "hep-th/0608150",
    archivePrefix = "arXiv",
    doi = "10.1007/s11005-006-0134-y",
    journal = "Lett. Math. Phys.",
    volume = "79",
    pages = "143--159",
    year = "2007"
}

@article{Basile:2019pic,
    author = "Basile, Thomas and Joung, Euihun and Park, Jeong-Hyuck",
    title = "{A note on Faddeev--Popov action for doubled-yet-gauged particle and graded Poisson geometry}",
    eprint = "1910.13120",
    archivePrefix = "arXiv",
    primaryClass = "hep-th",
    doi = "10.1007/JHEP02(2020)022",
    journal = "JHEP",
    volume = "02",
    pages = "022",
    year = "2020"
}

\end{document}